%% file: Pantheon_morphology.tex
\documentclass[twocolumn]{aastex63}

\shorttitle{Dependency of SN~Ia parameters on host morphology for \textsc{Pantheon}}
\shortauthors{Pruzhinskaya et al.}

\usepackage{graphicx}	
\usepackage{amsmath}	
\usepackage{amssymb}	
\usepackage{url}
\usepackage{longtable}
\usepackage{tabularx}
\usepackage{multirow}
\usepackage{booktabs}
\usepackage{natbib}
\usepackage{soul}

\begin{document}

\title{The dependency of Type Ia Supernova parameters on host galaxy morphology for the \textsc{Pantheon} cosmological sample}

\correspondingauthor{Maria Pruzhinskaya}
\email{pruzhinskaya@gmail.com}

\author{M.V.~Pruzhinskaya}
\affiliation{Lomonosov Moscow State University, Sternberg Astronomical Institute, Universitetsky pr.~13, Moscow, 119234, Russia}

\author{A.K.~Novinskaya}
\affiliation{Lomonosov Moscow State University, Sternberg Astronomical Institute, Universitetsky pr.~13, Moscow, 119234, Russia}
\affiliation{Lomonosov Moscow State University, Faculty of Physics, Leninskie Gory, 1-2, Moscow, 119991, Russia}

\author{N.~Pauna}
\affiliation{Universit\'{e} Clermont Auvergne, CNRS/IN2P3, LPC, Clermont-Ferrand, France}

\author{P.~Rosnet}
\affiliation{Universit\'{e} Clermont Auvergne, CNRS/IN2P3, LPC, Clermont-Ferrand, France}
 
\begin{abstract}
Type Ia Supernovae (SNe~Ia) are widely used to measure distances in the Universe. Despite the recent progress achieved in SN~Ia standardisation, the Hubble diagram still shows some remaining intrinsic dispersion. The remaining scatter in supernova luminosity could be due to the environmental effects that are not yet accounted for by the current standardisation methods. In this work we compare the local and global colour $(U-V)$, the local star formation rate, and the host stellar mass to the host galaxy morphology. The observed trends suggest that the host galaxy morphology is a good parameter to characterize the SN~Ia environment. Therefore, we study the influence of host galaxy morphology on light-curve parameters of SNe~Ia for the \textsc{Pantheon} cosmological supernova sample. We determine the Hubble morphological type of host galaxies for a sub-sample of 330 SNe~Ia. We confirm that the SALT2 stretch parameter $x_1$ depends on the host morphology with the $p$-value $\sim10^{-14}$. The supernovae with lower stretch value are hosted mainly by elliptical and lenticular galaxies. No correlation for the SALT2 colour parameter $c$ is found. We also examine Hubble diagram residuals for supernovae hosted by the ``Early-type''  and ``Late-type''  morphological groups of galaxies. The analysis reveals that the mean distance modulus residual in  early-type  galaxies  is  smaller  than the one in late-type galaxies, which means that early-type galaxies contain brighter supernovae.  However,  we do not observe any difference in the residual dispersion for these two morphological groups.

The obtained results are in the line with other analyses showing environmental dependence of SN~Ia light-curve parameters and luminosity. We confirm the importance of including a host galaxy parameter into the standardisation procedure of SNe~Ia for further cosmological studies.
\end{abstract}


\section{Introduction}
Type Ia Supernovae (SNe~Ia) stand out among the other types of supernovae in that they have smaller luminosity dispersion at maximum light and show higher optical luminosities. These two properties allowed to use them as cosmological distance indicators that led to the discovery of the accelerating expansion of the Universe~\citep{Riess98,Perlmutter99}. The most recent analysis of SNe~Ia indicates that considering the flat $\Lambda$CDM cosmology, the Universe is accelerating with $\Omega_\Lambda = 0.702\pm0.022$~\citep{2018ApJ...859..101S}.

When the first supernova light curves (LCs) had been collected and analysed, Walter Baade noticed that SNe are more uniform than novae, which makes them suitable as extragalactic distance indicators~\citep{1938ApJ....88..285B}. That time, Rudolph Minkowski has not yet divided SNe into two main types, Type~I and Type~II~\citep{Minkowski1941}. However, the idea that had been first expressed by Baade was confirmed later for Type Ia supernovae. It is how the ``standard candle'' hypothesis appeared. 

Now we know that the similarity of SN~Ia light curves and luminosities is explained by the similarity of the physical processes that lead to the outburst phenomenon. Generally, the outburst is a thermonuclear explosion of a C-O white dwarf whose mass has become close to or larger than the Chandrasekhar limit. In fact, when the detailed observations of a large number of supernovae had been accomplished, it became clear that the absolute magnitude at maximum can vary within $\sim1$~mag. The reasons of luminosity dispersion could be different. First, we are still uncertain about the nature of the progenitor systems of SNe~Ia. It can be the ``single-degenerate'' (SD) scenario where the burst is a result of the matter accretion on a white dwarf from a companion star~\citep{1973ApJ...186.1007W} or the ``double-degenerate'' (DD) scenario that is the merger of two white dwarfs~\citep{1984ApJS...54..335I,1984ApJ...277..355W}. To explain the peculiar Type Ia Supernovae (91bg-like, 91T-like, Iax) there exist some alternative scenarios, like sub-Chandrasekhar, that is usually associated with weak explosions, or super-Chandrasekhar scenario for more luminous events~\citep{2019ApJ...873...84P,2018A&A...618A.124F}. These scenarios have internal freedom that results in significant variations in observed light curves of SNe Ia: like point of deflagration-to-detonation transition (for SD scenario) or difference in total mass of merging white dwarfs (for DD scenario). 

Another important factor which could violate the ``standard candle'' hypothesis is dust. Dust around the supernovae, as well as in the host galaxy, surely affects light curve behaviour. The distribution and the properties of dust in host galaxies of supernovae could be different from that in the Milky Way. In the recent paper of~\cite{2020arXiv200410206B} it is also suggested that the dominant component of observed SN Ia intrinsic scatter is from $R_V$ variation of dust around a supernova.

In addition, the initial chemical composition of the progenitor stars also complicates the picture. A lower metallicity involves an increase of the Chandrasekhar limit. Indeed, according to~\cite{2011ARep...55..497B} the average energy of SNe Ia should increase from the redshift $z>2$ and increase significantly from the redshift $z>8$, since at the early stages of the Universe evolution more massive white dwarfs merged on average than now. However, so distant Type Ia Supernovae are not yet discovered.

Moreover, SNe~Ia explode in all types of galaxies that have an environment with different properties. In elliptical galaxies or in halo of spiral galaxies only old, i.e. metal-poor, stars with an age comparable to that of the Universe are located. On the contrary in the star formation regions of spiral galaxies there are young metal-rich stars. These factors (the age, the chemical composition of the region around a supernova, the presence of dust) could be considered as the environmental effects. 

Fortunately, it was established that supernovae are partly ``standardisable candles'' (see~Section~\ref{RP}), that allowed to improve a lot the accuracy of distance measurements and to reduce the intrinsic dispersion of SNe~Ia on the Hubble diagram  to 0.11~mag~\citep{Betoule2014,2018ApJ...859..101S}. A part of the remaining scatter in supernova luminosity could be due to the environmental effects that are not accounted by the current standardisation methods. Therefore, the SN~Ia standardisation procedure is one of the main sources of systematic uncertainties in the cosmological analyses.

In this paper we study how the host galaxy morphology affects the light-curve parameters of Type Ia SNe and therefore, their luminosity. The analysis is based on the most up-to-date cosmological sample of SNe~Ia, \textsc{Pantheon}~\citep{2018ApJ...859..101S}. The paper is organised as follows. In Section~\ref{EE} we describe the current supernova standardisation procedure and compare the different approaches to characterise the supernova environment. In Section~\ref{Data} we describe the \textsc{Pantheon} supernova sample and host morphological classification; we also show there how the host morphology affects the SN~Ia light-curve parameters and the Hubble diagram residuals. In Section~\ref{Discussion} we compare our results with the ones for other environmental parameters. Finally, we conclude this study in Section~\ref{Conclusions}.

\section{Environmental effects}
\label{EE}

\subsection{Supernova standardisation}
\label{RP}
The use of Type Ia Supernovae to measure the cosmological parameters of the Universe would never be possible without the discovery of the relation between the peak luminosity of SNe~Ia and their light curve decline rate after the maximum light. The relation was independently discovered by B.~W.~Rust and Yu.~P.~Pskovskii in the 1970s~\citep{1974PhDT.........7R,1977SvA....21..675P,1984SvA....28..658P}. It was also confirmed by M.~Phillips on a new level of accuracy using the better supernova sample~\citep{Phillips93}. The relation shows that the light curves of more luminous supernovae have slower decline rate after the maximum light. Later it has been found that SN~Ia absolute magnitude depends on the supernova colour as well ~\citep{1996AJ....112.2391H,Tripp98}. 

Nowadays more sophisticated parameters describing supernova observational properties are used to standardise SNe~Ia. Among the most recent models of SN~Ia parametrisation are \textsc{SALT2}~\citep{Guy07}, \textsc{SNEMO}~\citep{2018ApJ...869..167S}, and \textsc{SUGAR}~\citep{2020A&A...636A..46L}. 

To characterise the supernova LCs we use \textsc{SALT2} $x_1$ (stretch) and $c$ (colour) parameters. The $x_1$ parameter describes the time-stretching of the light curve. The $c$ parameter is the colour offset with respect to the average at the date of maximum luminosity in $B$-band, i.e. $c = (B - V )_{max} - \langle B - V\rangle$. We adopt the classical standardisation equation of the distance modulus:

\begin{equation}
\mu=m^*_B - M_B + \alpha x_1 - \beta c, 
\label{stdz}
\end{equation}
where $m^*_B$ --- value of the $B$-band apparent magnitude at maximum light, $M_B$ is a standardised absolute magnitude of the SNe~Ia in $B$-band for $x_1 = c = 0$; $\alpha$ and $\beta$ describe, consequently, the stretch and colour law for the whole SN Ia population.

\subsection{Local vs. global parameters}
\label{LocalGlobal}
The environment of SNe~Ia can be characterised by different parameters that we roughly divide into global and local. The global parameters are related to the whole host galaxy of supernova. It can be the host galaxy morphology, the metallicity, the stellar mass, the global colour, or the star formation rate (SFR). The local parameters in turn characterise the environment in a few kiloparsecs around a supernova, i.e. the local colour, the local SFR, the local specific SFR, etc. It is obvious that the local parameters provide more accurate description of the SN environment. However, the current state of the data processing and the resolution of the largest telescopes do not allow to measure the local parameters at high redshifts with a good accuracy or it becomes a very time-consuming process.  That is why a study of influence of the local parameters on the SNe~Ia properties is based mainly on the low-redshift supernova samples. For example, the most recent analysis of the local specific SFR in 1~kpc region around a supernova is done for 141 objects of the Nearby Supernova Factory~\citep{Aldering02} with redshift $0.02 < z < 0.08$~\citep{2018arXiv180603849R}. From that point it is more expedient to use the global parameters, for example, host galaxy morphology. At the moment, it is possible to determine the morphology of the most distant Hubble galaxies with $z > 1$~\citep{Meyers_2012}, which makes the study of host morphology impact possible even for cosmological supernovae.

Moreover, the number of discovered supernovae increases dramatically. In the epoch of the Legacy Survey of Space and Time (LSST;~\citealt{LSST2009}) millions of SNe will be discovered every year. In this sense the accurate measurements of the local environmental parameters for each supernova become very expensive since it requires time on the largest telescopes. The global parameters on the contrary are easier to obtain by processing the images of wide-field photometric surveys with use of traditional astronomical methods as well as machine learning techniques (e.g.~\citealt{10.1093/mnras/sty338}). 

\begin{figure}
    \includegraphics[width=\columnwidth]{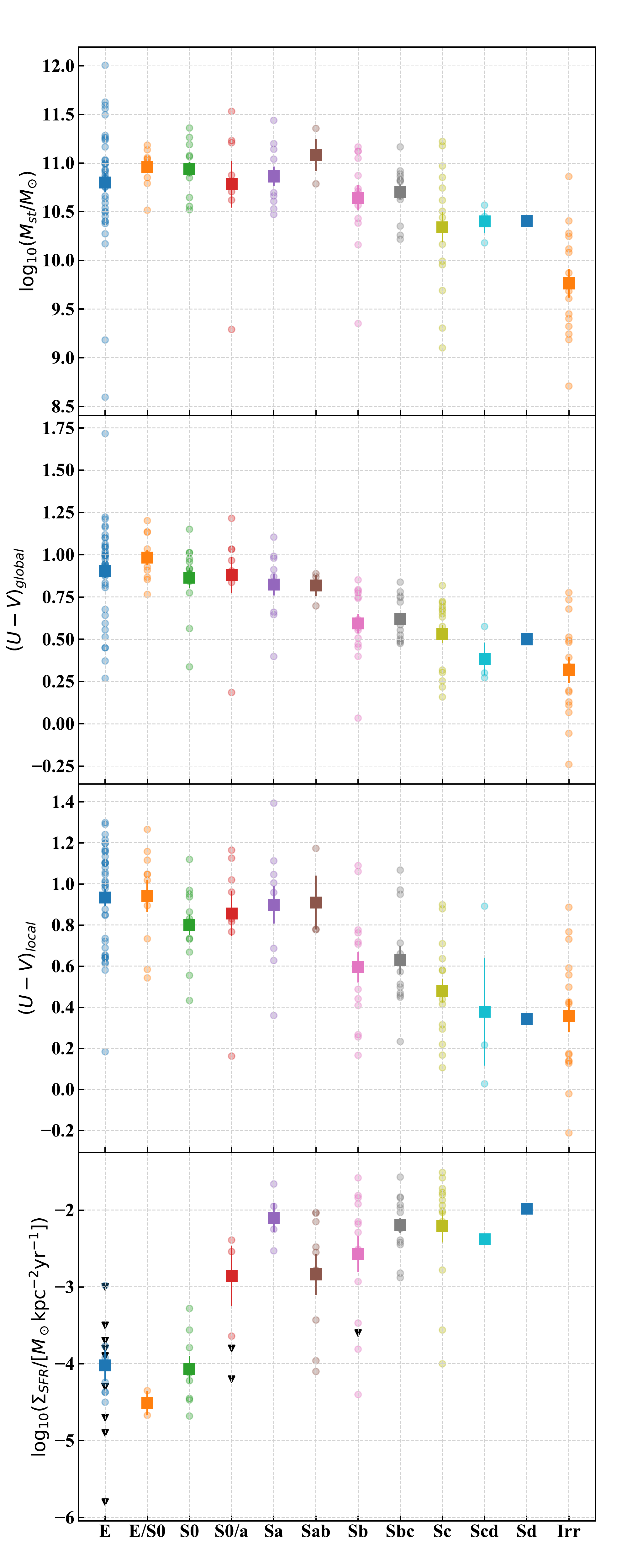}
    \caption{Stellar mass, global $(U-V)$ colour, local $(U-V)$ colour in 3-kpc region around SN Ia, and local star formation rate in 1~kpc region around SN Ia vs. morphological type of the supernova hosts for the SN sub-samples from~\citealt{2018A&A...615A..68R} (three upper plots) and from~\citealt{2015ApJ...802...20R} (lower plot). Mean values of the data points and associated standard deviations in each morphological bin are marked with squares with error bars. The upper limits on the local SFR are marked with triangles.}
    \label{fig:env}
\end{figure}

It is worth to stress that the local and global parameters correlate to each other. For example, the local $(U-V)$ rest-frame colour in a region of 3 kpc around a supernova correlates with the stellar mass of the host so that the most massive galaxies are those for which the close supernova environment is red (see figure 10 of~\citealt{2018A&A...615A..68R}). Here, we consider how the host morphology correlates with the local and other global parameters of environment. To do that we determine the supernova host morphology of 89 supernovae from~\cite{2015ApJ...802...20R} and 103 supernovae from~\cite{2018A&A...615A..68R} using SIMBAD\footnote{\url{http://simbad.u-strasbg.fr/simbad/}}~\citep{2000A&AS..143....9W}, HyperLEDA\footnote{\url{http://leda.univ-lyon1.fr/}}~\citep{2014A&A...570A..13M}, and NED\footnote{\url{https://ned.ipac.caltech.edu/}}~\citep{1988ESOC...28..335H,2007ASPC..376..153M} astronomical databases. To perform the comparison we use the local and global $(U-V)$ colour, the host galaxy stellar mass~\citep{2018A&A...615A..68R}, and the local star formation rate~\citep{2015ApJ...802...20R}. The results are given in Fig.~\ref{fig:env}. We observe the correlation between the host morphology and all considered parameters. Nevertheless, the morphological type dependency seems to be more pronounced with local parameters than with global ones.
 
To quantify the ability of host galaxy morphology to account for different mass, global or local colour, or star formation rate, we perform the Welch's $t$-test, or unequal variances $t$-test~\citep{10.1093/biomet/34.1-2.28,10.1093/beheco/ark016}. 
Generally speaking this is a two-sided test for the null hypothesis that two normally-distributed populations have equal means. Rather than the standard Student's $t$-test, Welch's $t$-test is more reliable when the two samples have unequal variances and/or unequal sample sizes. To perform the test we use the  \textsc{SciPy.stat} \textsc{Python} package\footnote{\url{https://docs.scipy.org/doc/scipy/reference/generated/scipy.stats.ttest_ind.html}}~\citep{2020SciPy-NMeth}. In this version of $t$-test, for two independent populations $n_1$ versus $n_2$ of means $\mu_1$ versus $\mu_2$ and standard deviations $s_1$ versus $s_2$, the $t$ variable supposed to follow the Student's probability law is built

\begin{equation}
t = \frac{\mu_1 - \mu_2}{\sqrt{\frac{s_1^2}{n_1}+\frac{s_2^2}{n_2}}},
\end{equation}
with a degree of freedom approximated to
\begin{equation}
\nu = \frac{\left(\frac{s_1^2}{n_1}+\frac{s_2^2}{n_2}\right)^2}{\frac{s_1^4}{(n_1-1)n_1^2}+\frac{s_2^4}{(n_2-1)n_2^2}}.
\end{equation}
Once $t$ and $\nu$ are calculated, the probability or $p$-value to obtain the null hypothesis is computed following the Student's $t$-distribution. The smaller $p$-value corresponds to higher separation of the two populations with respect to the variable under study or, in other words, the ability of morphology groups to account for different astrophysical properties of two populations.

The results of the $t$-test for the local and global parameters are reported in Table~\ref{tab-Ftest}. We split the data into two groups according to their morphological type. We also consider three different groupings based on the dependence observed in Fig.~\ref{fig:env}.
As can be seen from the Table~\ref{tab-Ftest}, the $p$-value varies from about $10^{-2}$ down to $10^{-12}$.
This quantitative test shows that depending on the considered parameter the optimal splitting into two morphological groups is not the same.
The SFR parameter is more powerful to separate  E--S0 group from S0/a--Irr, while the stellar mass and global and local $(U-V)$ colours are better to divide the galaxies into E--Sab and Sb--Irr groups. Therefore, this analysis suggests that the morphological type of a galaxy is a powerful parameter to separate the galaxy properties w.r.t. the colour and the star formation rate.
In conclusion, the grouping from E to S0/a morphology versus Sa to Irr is a good compromise to correlate both colours (local and global) and SFR with the two populations referred to below as ``Early-type'' (E--S0/a) and ``Late-type'' (Sa--Irr) morphological groups.

\begin{table*}
    \caption{\label{group} $p$-values of the Welch's $t$-test for the different morphological groupings corresponding to Fig.~\ref{fig:env} with respect to each global and local parameter.}
  \begin{tabular}{lcccc}
    \toprule
     Morph. group & $\log_{10}(M_{st}/M_\odot)$ & $(U-V)_{\rm global}$ & $(U-V)_{\rm local}$ & $\log_{10}(\Sigma_{\rm SFR} / [M_\odot \text{kpc}^{-2} \text{yr}^{-1}])$ \\
     \midrule
     E--S0 $|$ S0/a--Irr & $2.6 \times 10^{-2}$ & $6.2 \times 10^{-9}$ & $1.2 \times 10^{-6}$ & $1.7 \times 10^{-7}$ \\
     E--S0/a $|$ Sa--Irr & $1.7 \times 10^{-2}$ & $3.3 \times 10^{-10}$ & $3.4 \times 10^{-7}$ & $7.0 \times 10^{-7}$ \\
     E--Sab $|$ Sb--Irr & $6.6 \times 10^{-4}$ & $6.2 \times 10^{-13}$ & $2.4 \times 10^{-10}$ & $1.7 \times 10^{-4}$ \\
    \bottomrule
  \end{tabular}
  \label{tab-Ftest}
\end{table*}

Taking into account all of the above, in this work we use host galaxy morphology to describe the supernova environment and we study its impact on the \textsc{Pantheon}~\citep{2018ApJ...859..101S} cosmological sample of supernovae.

\section{Dependency of SN~Ia properties on host galaxy morphology}
In this section we examine the dependencies of the supernova light-curve parameters and luminosity on host morphology using SNe~Ia from the \textsc{Pantheon} sample. 
\label{Data}

\subsection{\textsc{Pantheon} supernova sample}
Cosmological supernova sample \textsc{Pantheon} consists of 1048\footnote{The exact number is 1047, since one supernova was counted twice under the different names, SN2005hj and SN6558.} spectroscopically confirmed SNe Ia with redshifts up to $z \simeq 2.3$~\citep{2018ApJ...859..101S}. \textsc{Pantheon} sample represents a compilation from several supernova surveys: 172 objects were taken from the nearby supernova surveys ($0.01 < z < 0.1$), 334 objects from the Sloan Digital Sky Survey (SDSS;~\citealt{2008AJ....135..338F,Kessler09}), 236 from the SuperNova Legacy Survey (SNLS;~\citealt{Guy10,Conley2011}), 279 objects from the Pan-STARRS survey (PS1;~\citealt{2014ApJ...795...44R,2014ApJ...795...45S}), and 26 SNe were discovered by the Hubble Space Telescope (HST;~\citealt{2004ApJ...607..665R,2007ApJ...659...98R,Suzuki2012,2014AJ....148...13R,2014ApJ...783...28G,Riess+2018}). \textsc{Pantheon} is the largest spectroscopic cosmological SN sample to date. The main advantages of \textsc{Pantheon} compared to the previous compilations are: an intercalibration between different surveys and a thorough investigation of systematic uncertainties.

\subsection{Morphological classification of host galaxies}
To analyse how the morphological type of host galaxy affects the supernova luminosity and standardisation parameters, we first determine the host morphology according to the Hubble morphological classification~\citep{1926ApJ....64..321H,1936rene.book.....H,1959HDP....53..275D}. To do that, we use SIMBAD, HyperLEDA, and NED astronomical databases as well as individual publications.

Unfortunately, it is not possible to find the detailed morphological classification for all supernova hosts, especially at high redshifts. For some supernovae we could only define either they belong to star-forming (SF) or passive (Pa) galaxies. For high-z SNe~Ia we use a classification from~\cite{2012ApJ...750....1M} and \cite{2014AJ....148...13R}. \cite{2012ApJ...750....1M} only distinguish passively evolving early-type galaxies from star-forming late-type galaxies. In~\cite{2014AJ....148...13R} SN hosts are classified visually into three main morphological categories (spheroid, disk, irregular) and two intermediate categories (spheroid+disk and disk+irregular). These morphological classes roughly correspond to broad bins over the Hubble sequence: spheroid (E/S0), spheroid+disk (S0/Sa), disk (Sb/Sbc/Sc), disk+irregular (Sc/Scd), irregular (Scd/Ir). It should be noticed that there are only few high-z HST supernovae and all of them, as well as their hosts, were subjected to the comprehensive astrophysical analysis in previous works. However, there are no such detailed studies for the host galaxies of SNLS supernovae. It explains the absence of morphological classification of supernova hosts at redshift $z\sim0.4-1$.

Based on these sources we found the host morphology of 330 SNe~Ia from the \textsc{Pantheon} sample. The result of this classification is given in Table~\ref{table_morph} (Appendix~\ref{appendix_1}). Columns 1 and 2 contain the supernova name and \textsc{Pantheon} ID, where 0 corresponds to low-z, 1 --- PS1, 2 --- SDSS, 3 --- SNLS, and 4 --- HST supernova sample. Column 3 contains the supernova redshift relative to the CMB frame. Host galaxy name is in column 4. The morphology extracted from SIMBAD, HyperLEDA and NED are given in columns 5, 6, 7, respectively. When the morphological classification provided by the different databases is controversial, we thoroughly analysed its primary source and defined a final type in column 8. In few cases the morphological classification is drawn out from the individual publications that we cite in column~9.   

\begin{figure*}
\begin{center}
\includegraphics[scale=0.6]{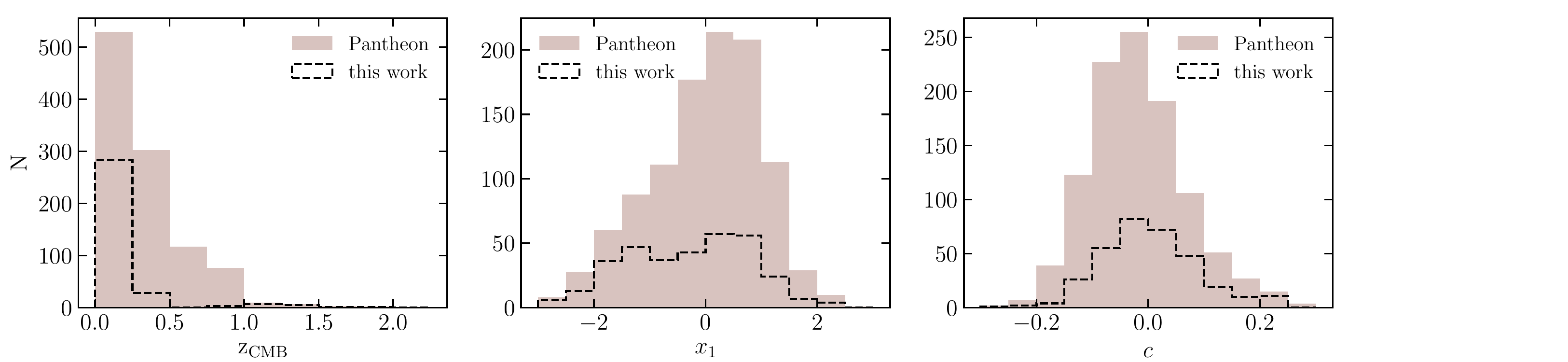}
\caption{Distribution of SNe~Ia by redshift ($z_{\rm CMB}$) and LC parameters such as stretch ($x_1$) and colour ($c$) for the whole \textsc{Pantheon} sample and for its sub-sample of 330 SNe~Ia used in this work.}
\label{hist}
\end{center}
\end{figure*}

In Fig.~\ref{hist} we show the distribution of 330 SNe with known host morphology by redshift, stretch, and colour parameters relative to the whole \textsc{Pantheon} supernova sample. The final distribution of SNe~Ia by host morphological type is summarised in Table~\ref{group}. 

\begin{table}
    \caption{\label{group} Distribution of the host galaxies of the \textsc{Pantheon} SN~Ia sub-sample according to their morphological type.}
\centering    
\begin{tabular}{@{}l|c@{}}
\toprule 
     Early-type (6) & \multirow{6}{*}{Early-type (91)}\\
     Pa (15)\\ 
     E (28)\\ 
     E/S0 (18)\\ 
     S0 (12)\\ 
     S0/a (12)\\
     \midrule
     Sa (21) & \multirow{12}{*}{Late-type (239)}\\
     Sab (16)\\ 
     Sb (37)\\ 
     Sbc (37)\\ 
     Sc (37)\\ 
     Sb/Sbc/Sc (1)\\
     Scd (3)\\ 
     Sd (1)\\ 
     Scd/Ir (1)\\
     Ir (30)\\ 
     SF (48)\\ 
     Late-type (7)\\ 
\bottomrule
\end{tabular}
\end{table}

\subsection{Results}
As we can see from Table~\ref{group} the distribution of the SN hosts by the morphological types is uneven. Moreover, while for the nearby galaxies the detailed Hubble classification is usually available, for the distant ones it is rather simplified. Therefore, for the further analysis we combine the ``close'' morphological types in two groups: ``Early-type'' and ``Late-type'' (see Table~\ref{group}). 
This classification in two groups is guided by the correlation observed between the host morphology and environmental parameters, as described in Section~\ref{LocalGlobal}. 
To the former group we assign all elliptical and lenticular galaxies as well as those classified as early-type or passive. From the environmental point of view these galaxies are dominated by the old, low-metallicity stars due to the low star formation rate. They are also relatively free from dust. The latter group is quite broad and includes all spirals, star-forming, late-type, and irregular galaxies. These systems contain the stars from different stellar populations and of different chemical composition. However, unlike early-type galaxies, they constantly form the new stars.

\subsection{$x_1$ and $c$ parameters}
\begin{figure*}
    \begin{minipage}{.49\textwidth}
        \centering
        \includegraphics[scale=0.39,trim={10mm 0 0 0},clip]{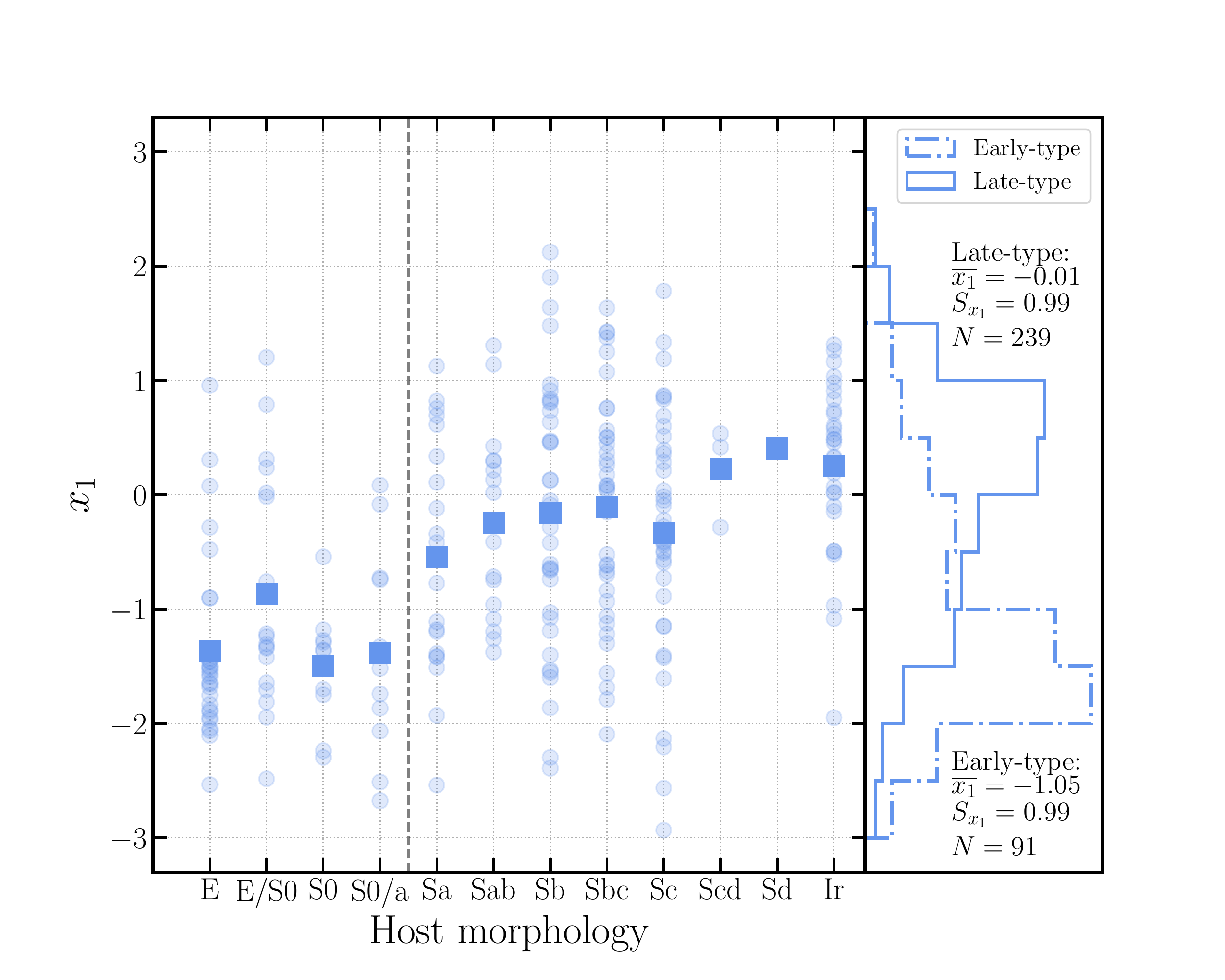}
    \end{minipage}
    \hfill
    \begin{minipage}{.49\textwidth}
        \centering
        \includegraphics[scale=0.39,trim={5mm 0 0 0},clip]{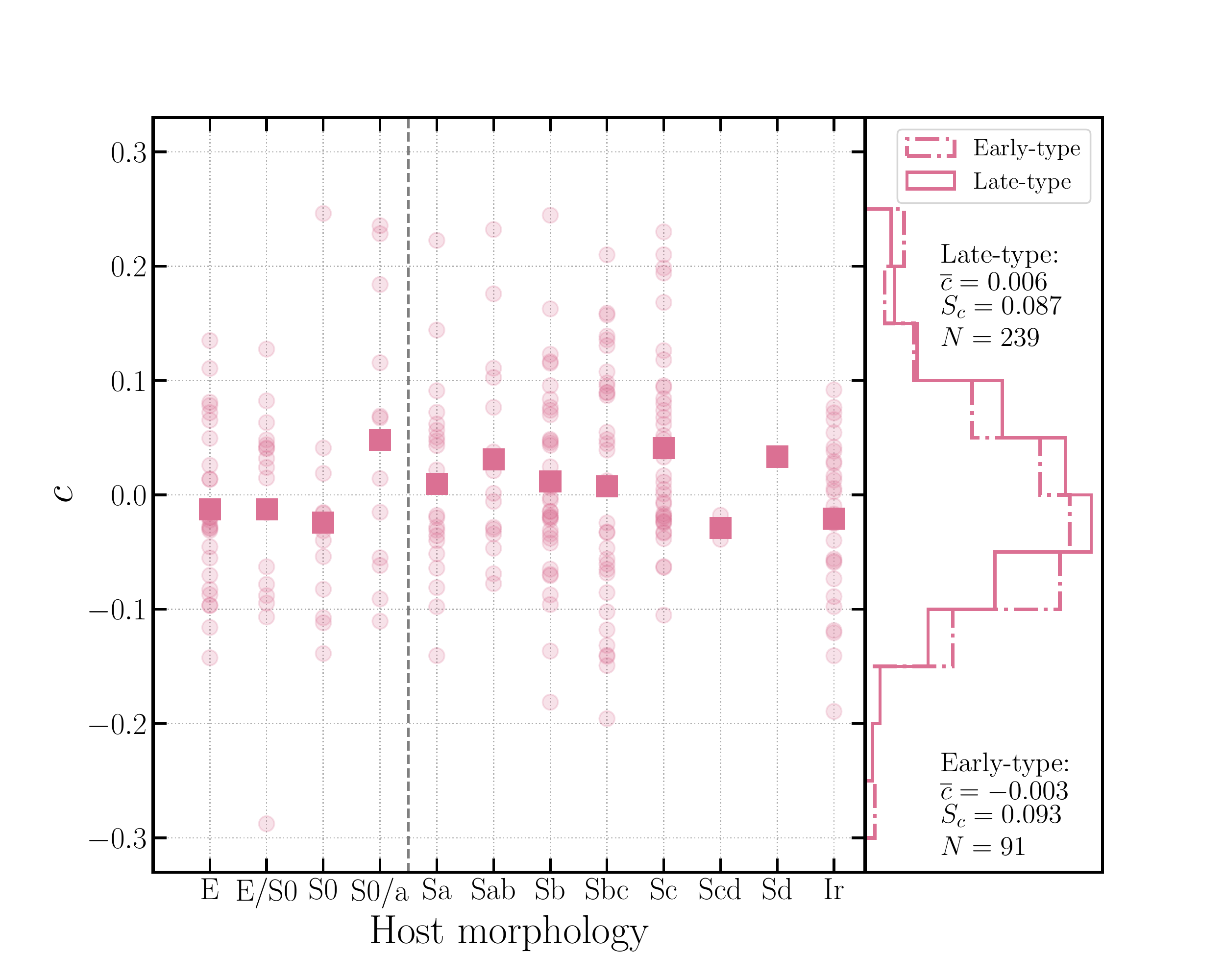}
    \end{minipage}  
    \caption{\textsc{SALT2} $x_1$ and $c$ light-curve parameters of SNe~Ia depending on the host morphology. The squares denote the mean values for the corresponding parameter in each morphological bin. The right subplots are the normalised histograms of $x_1$ and $c$ distributions for ``Early-type'' and ``Late-type'' morphological groups.}
        \label{fig:x1_c}
\end{figure*}

We first examine the dependency of SN Ia light-curve shape and colour parameters on host morphology. Fig.~\ref{fig:x1_c} shows the \textsc{SALT2} $x_1$ and $c$ light-curve parameters as a function of host galaxy morphology for the \textsc{Pantheon} SN~Ia sub-sample. For the left subplots we calculate the mean value of the corresponding LC parameter in each morphological bin. The mean values are marked with squares. The right subplots are the histograms of $x_1$ and $c$ distribution for the ``Early-type'' and ``Late-type'' morphological groups. As we are interested in the shape of the distribution, for clarity each histogram is normalised so that the integral under it equals one. 

We observe that the stretch parameter constantly increases along the Hubble morphological sequence from elliptical to irregular galaxies. If we consider only two morphological groups the difference in the stretch mean values is $\Delta_{\overline{x_1}} = 1.04$ with a significance $> 8.5 \sigma$ (Table~\ref{tab:x1_c}). Therefore, SNe Ia with the fastest decline rate, i.e. the most dim ones, appear in the galaxies with an older stellar population (elliptical and lenticular galaxies). The same conclusion is obtained by previous studies based on the other supernova samples~\citep{1995AJ....109....1H,1996AJ....112.2398H,2000AJ....120.1479H,1999AJ....117..707R,2003MNRAS.340.1057S,Henne2017,2019JKAS...52..181K}.

The difference in mean values for the colour parameter $c$ is observed neither for detailed morphological classification nor for two morphological groups. This is consistent with the previous results obtained by~\cite{2010MNRAS.406..782S,2019JKAS...52..181K}. \cite{Henne2017} found that SNe~Ia in elliptical and lenticular galaxies have slightly bluer colour than others, and explained this by the fact that the spiral galaxies contain more dust which makes the supernovae redder. However, the found difference was not statistically significant.

To check the significance of the results we perform the Welch's $t$-test described in Section~\ref{LocalGlobal}. The test confirmed that for SNe~Ia exploded in the ``Early-type'' and ``Late-type'' morphological groups the difference in $\overline{x_1}$ is significant with the $p$-value equal to $\sim10^{-14}$. On the other hand, the $p$-value of the colour parameter is 0.45 which is not significant (see Table~\ref{tab:x1_c}).

\begin{table*}
    \caption{Mean and standard deviation of the \textsc{SALT2} $x_1$ and $c$ light-curve parameters and the Hubble residuals $\Delta{\mu}$ for ``Early-type'' and ``Late-type'' morphological groups. Last row contains the $p$-values of the Welch's $t$-test used to compare the equality of the means.}
  \begin{tabular}{lccccccc}
    \toprule
     Morph. group & $N$ & $\overline{x_1}$ & $S_{x_1}$ & $\overline{c}$ & $S_c$ & $\overline{\Delta\mu}$ & $S_{\Delta\mu}$ \\
     \midrule
     Early-type & 91 & $-1.05\pm0.10$ & $0.99$ & $-0.003\pm0.010$ & $0.093$ & $-0.092\pm0.016$& $0.150$\\
     Late-type & 239 & $-0.01\pm0.07$ & $0.99$ & $0.006\pm0.006$ & $0.087$ & $-0.034\pm0.010$& $0.152$ \\
     \midrule
     $p$-value & & $8.8\times10^{-15}$ & & 0.45 & & $2.0 \times 10^{-3}$  & \\
    \bottomrule
  \end{tabular}
  \label{tab:x1_c}
\end{table*}

\subsection{Hubble residuals}
\begin{figure*}
\begin{center}
\includegraphics[scale=1,trim={0 35mm 0 0},clip]{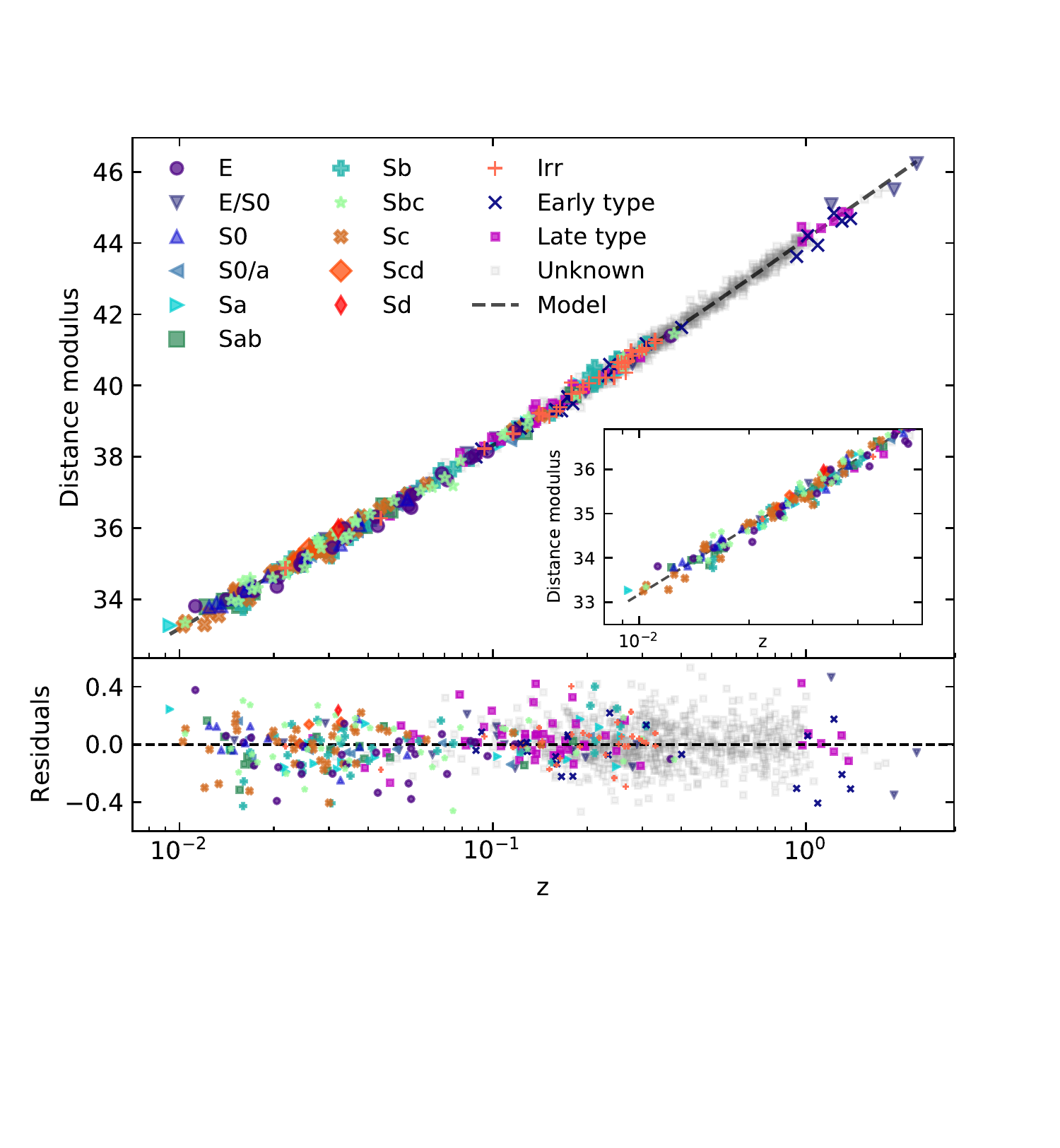}
\caption{Hubble diagram for the \textsc{Pantheon} supernovae. Different markers correspond to supernovae belonging to galaxies of different morphological types. The model corresponds to the flat $\Lambda$CDM cosmology with $\Omega_{\Lambda}=0.702\pm0.022$~\citep{2018ApJ...859..101S}.}
\label{HD}
\end{center}
\end{figure*}

To investigate whether Type Ia Supernovae can be physically different in the separate groups due to environmental effect, we reproduce the Hubble diagram from the \textsc{Pantheon} analysis. We consider the flat $\Lambda$CDM-model in which the Universe is filled with the matter (cold dark matter and ordinary matter) and the dark energy, i.e. $\Omega_m + \Omega_{\Lambda} = 1$. In this model, the theoretical distance modulus is given by

\begin{equation}
\mu_{\rm model} = 5\log_{10} d_{\rm L} - 5,
\end{equation}

\begin{equation}
d_{\rm L} = \frac{c}{H_0}(1+z)\int_{0}^{z} \frac{dz'}{\sqrt{\Omega_\Lambda + \Omega_m(1 + z')^3}},
\end{equation}
where $d_L$ is the luminosity distance. We assume $\Omega_{\Lambda}=0.702\pm0.022$~\citep{2018ApJ...859..101S}. The Hubble diagram is given in Fig.~\ref{HD}. It can be noticed, for example, that the HST supernovae from the early-type hosts lie below the ones exploded in the late-type galaxies. 

The observational distance modulus from \cite{2018ApJ...859..101S} contains a distance correction based on the supernova host galaxy mass~(see also the mass step introduced in~\citealt{Betoule2014}). The correction takes into account the correlation between host stellar mass and Hubble residuals, i.e. it is responsible for the environmental correction in the cosmological analyses. Therefore, to study the host morphology impact on the Hubble residuals we removed this correction from the observational distance modulus.

\begin{figure}
\begin{center}
\includegraphics[width=\columnwidth,trim={8mm 0 20mm 0},clip]{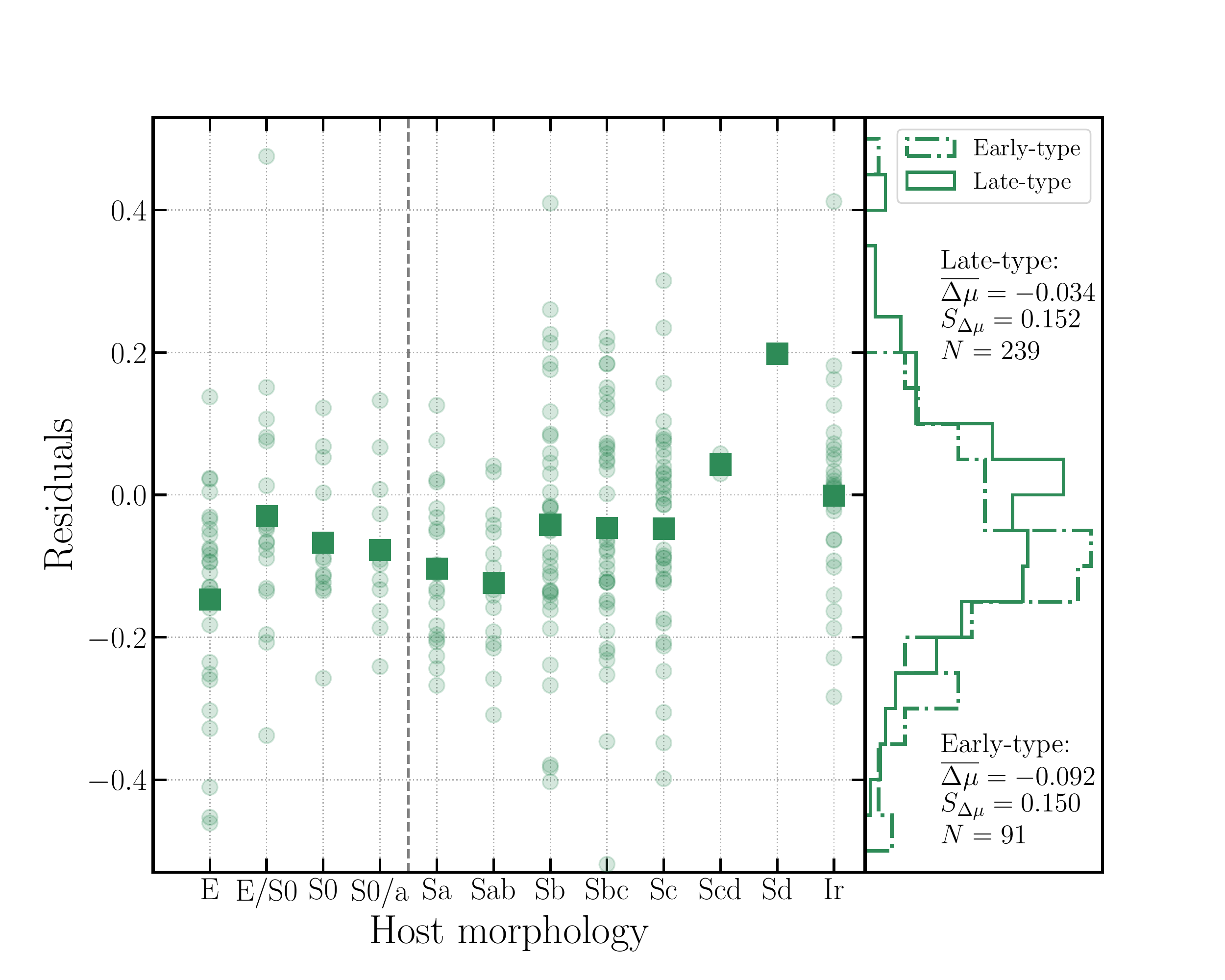}
\caption{Hubble diagram residuals of SNe~Ia depending on the host morphology. The squares denote the mean values $\Delta\mu$ in each morphological bin. The right subplot is the normalised histogram of $\Delta\mu$ distribution for ``Early-type'' and ``Late-type'' morphological groups.}
\label{fig:mu_res}
\end{center}
\end{figure}

The results are given in Fig.~\ref{fig:mu_res} and Table~\ref{tab:x1_c}. While from Fig.~\ref{fig:mu_res} it is not very clear how the residuals change along the Hubble sequence, if we divide the hosts into two morphological groups, we will see that the mean residual in the early-type galaxies is smaller than the one in the late-type galaxies. Therefore, SNe~Ia in the early-type hosts are brighter after the light-curve corrections than those in the late-type. According to the Welch's $t$-test this difference is significant with the $p$-value equal to $0.002$. The same result is found in~\cite{Henne2017}, however~\cite{2019JKAS...52..181K} do not observe any conclusive trend for the low-z and SDSS supernova samples.

\begin{figure}
\begin{center}
\includegraphics[width=\columnwidth,trim={8mm 0 20mm 0},clip]{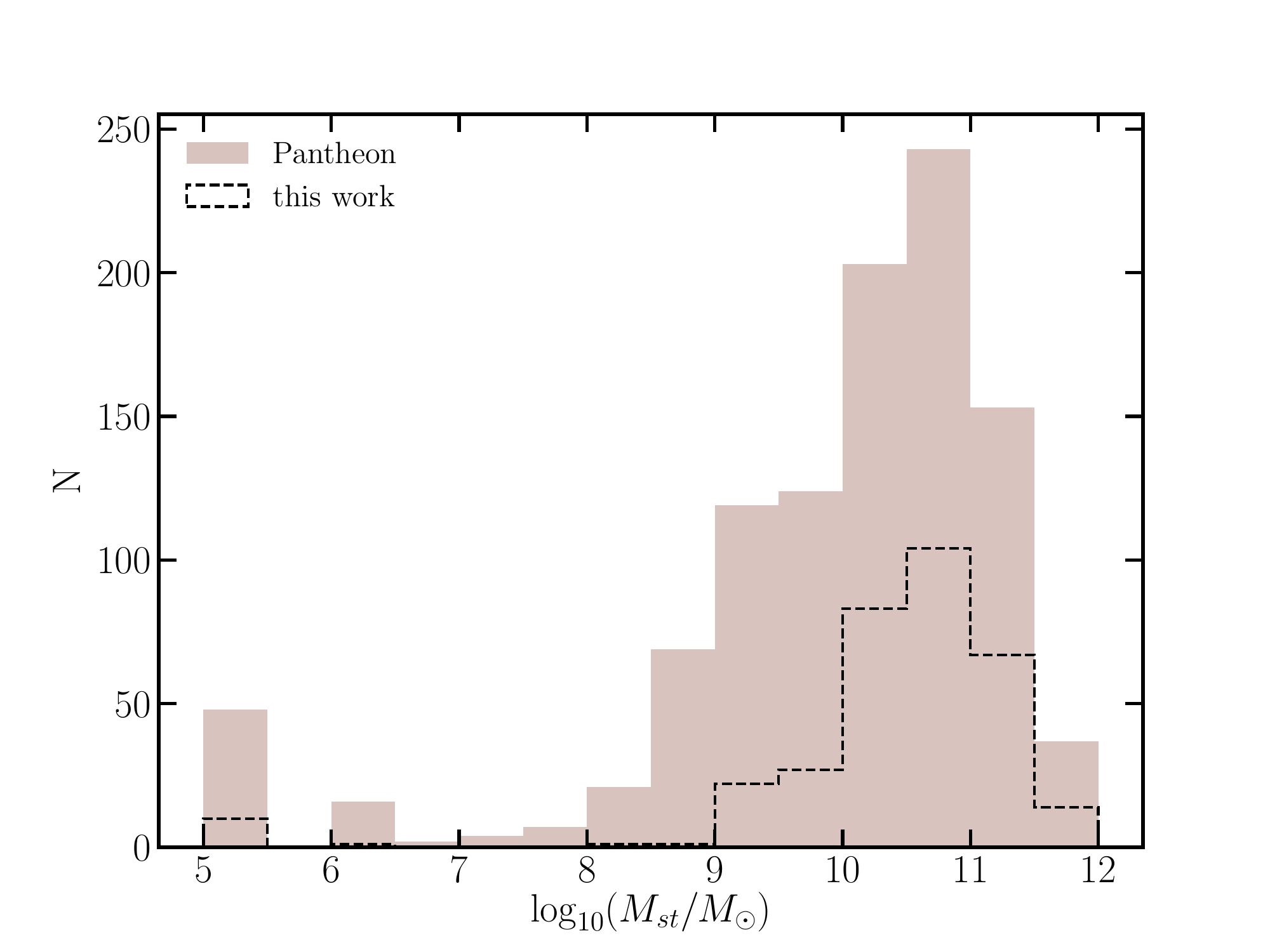}
\caption{Distribution of SNe Ia by the host stellar mass for the whole \textsc{Pantheon} sample and its sub-sample of 330 SNe Ia used in this work.}
\label{fig:hist_mst}
\end{center}
\end{figure}

It can be noticed that the residuals in Fig.~\ref{fig:mu_res} are mainly negative. To explain this, we plot the distribution of the SNe Ia sub-sample considered in this work by the host stellar mass (Fig.~\ref{fig:hist_mst}). For the majority of our sample $\log_{10} (M_{st}/M_\odot) > 10$. Meanwhile, the figure 14 of \cite{2018ApJ...859..101S} shows that the mean residuals for the \textsc{Pantheon} SNe with $\log_{10} (M_{st}/M_\odot) > 10$ are negative. Since galaxies with larger stellar mass are supposed to be more luminous, it is reasonable to suggest that it was easier to determine the morphological types of those ones than for the low-mass dim galaxies. Therefore, this can be a selection effect.

The host galaxy morphology could also affect the residual dispersion on the Hubble diagram. Our initial assumption is that SNe~Ia should be more homogeneous in the early-type galaxies due to the similar explosion mechanism and small dust contamination~\citep{2011NewA...16..250L,2011AstL...37..663P}. However, we do not see any difference in the residual dispersion for early-type and late-type hosts. Moreover, some previous studies show that SNe Ia in late-type spirals (Scd-Ir) are more homogeneous~\citep{Henne2017,2019JKAS...52..181K}.

\section{Discussion}
\label{Discussion}

\subsection{Comparison with the results for other environmental parameters}
In Section~\ref{LocalGlobal} we show that the different parameters of environment correlate with the host morphology. Indeed, previous studies mention that the low-stretch supernovae are preferentially hosted by the galaxies with little or no ongoing star formation that is consistent with our results for the early-type galaxies (e.g.~\citealt{2006ApJ...648..868S,2009ApJ...707.1449N,2010MNRAS.406..782S,2012ApJ...755...61S,2013MNRAS.435.1680J,2019JKAS...52..181K}). Moreover, the analyses of \cite{2009ApJ...707.1449N,2010MNRAS.406..782S,ChildressHostResidual,2013MNRAS.435.1680J,2019JKAS...52..181K} have revealed that the observed brightness of supernovae correlates with the host stellar mass, such that the more massive hosts produce mainly fast-decline rate (low-stretch) SNe~Ia. This result is consistent with ours, as illustrated by Fig.~\ref{fig:env} showing that the early-type hosts have the highest stellar mass on average. 

The relation between the colour parameter $c$ and the different host properties is less evident. While we do not see any connection between $c$ and host morphology, \citet{2010MNRAS.406..782S} claim that SNLS SNe~Ia in low specific SFR systems do show slightly bluer colours in the mean and find no difference in SN colours in low-mass and high-mass hosts. On the contrary,~\citet{2019JKAS...52..181K} do not observe any trend with global specific SFR but show that SNe~Ia in high-mass hosts are somewhat bluer than those in low-mass hosts. Moreover, \citet{ChildressHostResidual} notice that red SNe~Ia occur in high-metallicity galaxies. It is expected that the high-metallicity star-forming galaxies contain more dust that, therefore, should affect SN~Ia colours.

Finally, previous studies show that galaxies with higher star formation rate host on average fainter supernovae which is consistent with our results for the late-type (star-forming) galaxies (e.g.~\citealt{2010MNRAS.406..782S,2015ApJ...812...31J,2019JKAS...52..181K}).

\subsection{Perspectives}
The underlying motivation to use the host morphology as environmental parameter is that in contrast to the late-type galaxies, the elliptical galaxies are dominated by the old stellar population and contain the small amount of dust. However, in such approach we ignore the fact that in halo of spiral galaxies the conditions are very similar to those in elliptical ones. Thus, in the further work it seems promising to combine the morphological criteria with the information about the distance between the host center and the supernova position. Indeed, in \cite{2018MNRAS.481.2766H} the projected galactocentric distance to the host for a sub-set of the SDSS SNe~Ia has been investigated. It was shown that the scatter around the Hubble diagram is less for the SNe~Ia with larger galactocentric distances, i.e. they are more homogeneous. Due to the small statics, the significance of this result is only $1.4\sigma$, however it will be interesting to study this effect for a lager sample size and in combination with the knowledge of the host morphology. 

Since there is a significant difference in the stretch parameter for ``Early-type'' and ``Late-type'' morphological groups, we also expect a difference in $\alpha$ nuisance parameter from Eq.~\ref{stdz}. In other words, the standardisation of the SN~Ia luminosity variations in old environment is not the same as in young, star-forming environment (e.g.~\citealt{Henne2017}). 
Therefore, instead of adding a correction term to the standardisation Eq.~\ref{stdz}, we could also adapt the nuisance parameter $\alpha$ to the SN Ia environment. For instance, two $\alpha$ parameters, accounting for the different morphological groups defined in this work, could be used for the Hubble diagram fit. In this way, the difference in the stretch distribution will be accounted automatically in the cosmological fit. This new approach of the SN Ia environmental correction will be tested in a coming work.

\section{Conclusions}
\label{Conclusions}
In this paper, we studied the dependencies of the different attributes of SN~Ia environment, such as local and global colour $(U-V)$, local SFR, and stellar mass on host galaxy morphology in order to test the reliability of the morphology as a leverage environmental parameter. We found a significant correlation of the considered parameters with the host morphology and confirmed its ability to describe the properties of the supernova host galaxies.

Then, we studied the influence of host galaxy morphology on the supernova light-curve parameters. We believe that host morphology can be a good environmental parameter for several reasons. First, it is possible that a SN explosion depends on the chemical composition of the progenitor. The elliptical galaxies contain mainly the oldest, first-generation metal-poor stars, which leads to a more homogeneous chemical composition of SN progenitors. Then, there are several progenitor scenarios that could lead to the different supernova luminosity and its LC parameters. We expect that SNe~Ia in elliptical galaxies explode via the double degenerate scenario. At last, the dust properties matter. Elliptical galaxies are relatively dust-free. The role of the above listed factors is difficult to evaluate in the theoretical studies, although some progress is achieved~(e.g.~\citealt{Umeda_1999,2003ApJ...590L..83T,2009Natur.460..869K}).

Using the astronomical databases and the individual publications, we determined the Hubble morphological type of host galaxies of 330 \textsc{Pantheon} SNe~Ia. We confirmed that the SALT2 stretch parameter $x_1$ is correlated with the host galaxy type. The supernovae with a lower stretch value are hosted mainly by elliptical and lenticular galaxies. The correlation for the SALT2 colour parameter $c$ has not been found. The analysis revealed that the mean distance modulus residual $\overline{\Delta\mu}$ in early-type galaxies is smaller than the one in late-type galaxies, which means that early-type galaxies host  brighter supernovae. However, we did not see any difference in the residual dispersion for these two morphological groups. Our results for the stretch parameter and the residual values are consistent with the previous works. The conclusions concerning the colour parameter and residual dispersion are less evident since the results of the previous studies are dependent on the choice of the environmental parameter and of the supernova sample (see Section~\ref{Discussion}).

Therefore, we confirm the variation of the light curve parameters, as well as the Hubble residuals, with morphological type of host galaxy. The including a host galaxy parameter into the SN~Ia standardisation and the Hubble diagram fit is expected to be important for further cosmological studies.

\section*{Acknowledgements}
M.V.P. and A.K.N. acknowledge support from RSF grant 18-72-00159 for the analysis of the environmental effects for the \textsc{Pantheon} supernova sample. M.V.P. and A.K.N. acknowledge the Program of Development of M.V.~Lomonosov Moscow State University (Leading Scientific School ``Physics of stars, relativistic objects and galaxies''). N.P. and P.R. acknowledge the University Clermont Auvergne and the French CNRS-IN2P3 agency for their funding support.

This research has made use of the SIMBAD database,
operated at CDS, Strasbourg, France. We acknowledge the usage of the HyperLeda database (\url{http://leda.univ-lyon1.fr}). This research has made use of the NASA/IPAC Extragalactic Database (NED), which is funded by the National Aeronautics and Space Administration and operated by the California Institute of Technology. This research has made use of NASA's Astrophysics Data System Bibliographic Services and following {\sc Python} software packages: {\sc Matplotlib}~\citep{matplotlib}, {\sc NumPy}~\citep{numpy}, {\sc SciPy}~\citep{2020SciPy-NMeth}, {\sc pandas}~\citep{mckinney2010data,mckinney2011pandas}.

\bibliographystyle{aasjournal}
\bibliography{Biblio}

\appendix

\section{Supplementary Data}
\label{appendix_1}
\input{online_mat.tex}

\end{document}

%% file: online_mat.tex
{\footnotesize 
\begin{longtable}{lllllllll} 
\caption{Host galaxy morphology of Type Ia Supernovae from the \textsc{Pantheon} sample~\citep{2018ApJ...859..101S} found in SIMBAD ([I],~\citealt{2000A&AS..143....9W}), HyperLEDA ([II],~\citealt{2014A&A...570A..13M}), NED ([III],~\citealt{1988ESOC...28..335H,2007ASPC..376..153M})  databases or individual publications cited in the column ``Reference''. The final type is summarised in column ``Type''. Object ID denotes the supernova survey included in \textsc{Pantheon}: 0 --- low-z, 1 --- PS1, 2 --- SDSS, 3 --- SNLS, 4 --- HST supernovae.} 
\label{table_morph}\\ 
\hline
SN & ID & $z_{\rm CMB}$ & Host galaxy & [I] & [II] & [III] & Type & Reference \\
\hline
2009an	&	0	&	0.00931	&	NGC 4332	&	Sa	&	Sa	&	Sa	&	Sa	&		\\
2002cr	&	0	&	0.01025	&	NGC 5468	&	Sc	&	Sc	&	Scd 	&	Sc	&		\\
2006bh	&	0	&	0.01042	&	NGC 7329	&	Sc:	&	Sbc	&	Sb	&	Sbc	&		\\
2002dp	&	0	&	0.01045	&	NGC 7678	&	Sbc	&	Sc	&	Sc	&	Sc	&		\\
2010Y	&	0	&	0.01123	&	NGC 3392	&	E	&	E	&	E?	&	E	&		\\
1998dk	&	0	&	0.01202	&	UGC 139	&	Scd	&	Sc	&	Sc? 	&	Sc	&		\\
2002ha	&	0	&	0.01224	&	NGC 6962	&	Sab	&	Sa	&	Sab	&	Sab	&		\\
2009kk	&	0	&	0.01243	&	2MASX J03494330-0315348	&		&		&		&	S0	&	\citet{2016AJ....152...50T}	\\
2009kq	&	0	&	0.01247	&	MCG+05-21-01	&	Sbc	&	Sc	&		&	Sc	&		\\
1997E	&	0	&	0.01313	&	NGC 2258	&	S0 	&	S0	&	S0	&	S0	&		\\
1999dq	&	0	&	0.01334	&	NGC 976	&	Sc	&	Sbc	&	Sc:	&	Sc	&		\\
2008hv	&	0	&	0.01359	&	NGC 2765	&	S0	&	S0	&	S0	&	S0	&		\\
2005kc	&	0	&	0.01390	&	NGC 7311	&	Sa	&	Sab	&	Sab	&	Sab	&		\\
2006N	&	0	&	0.01408	&	MCG+11-08-012	&		&	E	&		&	E	&		\\
2001fe	&	0	&	0.01449	&	UGC 5129	&	Sa	&	Sa	&	Sa	&	Sa	&		\\
2004eo	&	0	&	0.01457	&	NGC 6928	&	Sab	&	Sab	&	Sab	&	Sab	&		\\
2004ey	&	0	&	0.01462	&	UGC 11816	&	Sbc	&	SBc	&	Sc:	&	Sbc	&		\\
2005el	&	0	&	0.01489	&	NGC 1819	&	S0	&	S0	&	S0	&	S0	&		\\
2006hb	&	0	&	0.01496	&	MCG-04-12-034	&	E/S0	&	E-S0	&	E?	&	E/S0	&		\\
2006td	&	0	&	0.01504	&	2MASX J01581578+3620538	&	S	&	Sc	&		&	Sc	&		\\
2007ca	&	0	&	0.01515	&	MCG-02-34-61	&	Sc	&	Sc	&	Sc	&	Sc	&		\\
2009nq	&	0	&	0.01526	&	NGC 7549	&	Sbc	&	Sc	&	Scd	&	Sc	&		\\
1999ej	&	0	&	0.01544	&	NGC 495	&	S0a	&	S0-a	&	S0/a	&	S0/a	&		\\
2001en	&	0	&	0.01544	&	NGC 523	&	Sb	&	Sbc	&		&	Sbc	&		\\
2005bo	&	0	&	0.01556	&	NGC 4708	&	Sab	&	Sa	&	Sab	&	Sab	&		\\
2007A	&	0	&	0.01595	&	NGC 105	&	Sbc	&	Sab	&		&	Sbc	&		\\
2001V	&	0	&	0.01596	&	NGC 3987	&	Sb	&	Sb	&	Sb	&	Sb	&		\\
2000dk	&	0	&	0.01602	&	NGC 382	&	E:	&	E	&	E:	&	E	&		\\
1998ef	&	0	&	0.01602	&	UGC 646	&	S	&	Sb	&	S?	&	Sb	&		\\
1994S	&	0	&	0.01611	&	NGC 4495	&	E	&	Sab	&	Sab	&	Sab	&		\\
2010H	&	0	&	0.01621	&	IC 494	&	S0	&	S0	&	S0:	&	S0	&		\\
2001da	&	0	&	0.01647	&	NGC 7780 	&	Sab	&	Sa	&	Sab	&	Sab	&		\\
2007ap	&	0	&	0.01668	&	MCG+03-41-003	&	S0	&	S0-a	&	S0	&	S0/a	&		\\
1996bv	&	0	&	0.01673	&	UGC 3432	&	Sc	&	Sc	&	Scd:	&	Sc	&		\\
1997Y	&	0	&	0.01678	&	NGC 4675	&	Sb	&	Sb	&	Sb:	&	Sb	&		\\
2007fb	&	0	&	0.01681	&	UGC 12859	&	Sbc	&	Sbc	&	Sbc	&	Sbc	&		\\
2006ef	&	0	&	0.01682	&	NGC 809	&	S0	&	S0	&	S0:	&	S0	&		\\
1993ae	&	0	&	0.01693	&	UGC 1071	&	E	&		&	S?	&	E	&		\\
2009le	&	0	&	0.01703	&	2MASX J02091807-2324542	&	Sc	&	Sbc	&	Sbc	&	Sbc	&		\\
2001G	&	0	&	0.01707	&	MCG+08-17-043	&		&	Sab	&	Sc	&	Sab	&		\\
2008C	&	0	&	0.01708	&	UGC 3611	&	S0a	&	S0-a	&	S0/a	&	S0/a	&		\\
2008L	&	0	&	0.01730	&	NGC 1259	&	E	&	E-S0	&		&	E	&		\\
2006ax	&	0	&	0.01773	&	NGC 3663	&	Sb	&	Sbc	&	Sbc	&	Sbc	&		\\
2006ej	&	0	&	0.01916	&	IC 1563	&	S0	&	S0	&	S0	&	S0	&		\\
2002kf	&	0	&	0.01948	&	2MASX J06371661+4951005	&		&		&		&	Sc	&	\citet{2016AJ....152...50T}	\\
2010A	&	0	&	0.01985	&	UGC 2019	&	I…	&	Sbc	&	S?	&	Sbc	&		\\
2008ds	&	0	&	0.01994	&	UGC 299	&	Sc	&	Sc	&	Sc	&	Sc	&		\\
1998ec	&	0	&	0.02010	&	UGC 3576	&	Sb	&	Sb	&	Sb	&	Sb	&		\\
2000B	&	0	&	0.02045	&	NGC 2320	&	E	&	E	&	E	&	E	&		\\
2009ds	&	0	&	0.02050	&	NGC 3905	&	Sc	&	Sc	&	Sc	&	Sc	&		\\
2005ki	&	0	&	0.02066	&	NGC 3332	&	E	&	E-S0	&	S0	&	E	&		\\
2006bq	&	0	&	0.02146	&	NGC 6685	&	E/S0	&	E-S0	&	S0:	&	E/S0	&		\\
2006et	&	0	&	0.02160	&	NGC 232	&	Sa	&	Sa	&	Sa?	&	Sa	&		\\
2006or	&	0	&	0.02167	&	NGC 3891	&	Sc	&	Sbc	&	Sbc	&	Sbc	&		\\
2000fa	&	0	&	0.02180	&	UGC 3770	&	I	&	I	&	Im	&	Ir	&		\\
2007bc	&	0	&	0.02187	&	UGC 6332	&	Sab	&	Sa	&	Sa	&	Sa	&		\\
1995ak	&	0	&	0.02193	&	IC 1844	&	Sbc	&	Sbc	&		&	Sbc	&		\\
2009na	&	0	&	0.02212	&	UGC 5884	&	Sc	&	Sb	&	Sb:	&	Sb	&		\\
2006mp	&	0	&	0.02280	&	MCG+08-31-029	&		&		&		&	Sb	&	\citet{2016AJ....152...50T}	\\
2006sr	&	0	&	0.02298	&	UGC 14	&	Sc	&	Sc	&	S?	&	Sc	&		\\
2000cn	&	0	&	0.02321	&	UGC 11064	&	Sc	&	Sc	&	Scd:	&	Sc	&		\\
2006cp	&	0	&	0.02334	&	UGC 7357	&	Sd	&	Sc	&	Sc	&	Sc	&		\\
1998eg	&	0	&	0.02362	&	MCG+01-57-014	&	Sc	&	Sc	&	Scd:	&	Sc	&		\\
2006ac	&	0	&	0.02395	&	NGC 4619	&	Sc	&	Sb	&	Sb	&	Sb	&		\\
2003it	&	0	&	0.02419	&	UGC 40	&	S	&	Sb	&	S?	&	Sb	&		\\
2007F	&	0	&	0.02419	&	UGC 8162	&	Scd	&	Sc	&	Scd:	&	Scd	&		\\
1994M	&	0	&	0.02431	&	NGC 4493	&	S0	&	E	&	E	&	E	&		\\
2008bf	&	0	&	0.02453	&	NGC 4055	&	E:	&	E	&	E:	&	E	&		\\
2009D	&	0	&	0.02466	&	MCG-03-10-52	&	Sb	&	Sb	&	Sb	&	Sb	&		\\
2002bf	&	0	&	0.02474	&	2MASX J10154226+5540030	&	Sbc	&	Sb	&	Sb:	&	Sb	&		\\
2002he	&	0	&	0.02484	&	UGC 4322	&	E	&	E	&	E	&	E	&		\\
2007cq	&	0	&	0.02510	&	2MASX J22144070+0504435	&		&		&		&	Sbc	&	\citet{Hakobyan+2012}	\\
2006bb	&	0	&	0.02524	&	UGC 4468	&	S0	&	S0	&	S0	&	S0	&		\\
2005M	&	0	&	0.02562	&	NGC 2930	&		&	Sbc	&	S?	&	Sbc	&		\\
1999X	&	0	&	0.02577	&	2MASX J08543185+3630346	&		&	Sa	&	Sa	&	Sa	&		\\
2005ms	&	0	&	0.02590	&	UGC 4614	&	Sd	&	Sb	&	S?	&	Scd	&	\citet{Hakobyan+2012}	\\
2005mc	&	0	&	0.02602	&	UGC 4414	&		&	S0-a	&	S0/a 	&	S0/a	&		\\
370356	&	1	&	0.02640	&	UGC 7228	&	Sab	&	Sb	&		&	Sb	&		\\
2007co	&	0	&	0.02656	&	MCG+05-43-016	&		&		&		&	Sc	&	\citet{2016AJ....152...50T}	\\
2007su	&	0	&	0.02662	&	LEDA 3374128	&		&		&		&	SF	&	\citet{Hakobyan+2012}	\\
2001gb	&	0	&	0.02676	&	IC 582	&	Sd	&	Sb	&	S	&	Sc	&	\citet{Hakobyan+2012}	\\
2005na	&	0	&	0.02683	&	UGC 3634	&	Sa	&	Sa	&	Sa	&	Sa	&		\\
2008ar	&	0	&	0.02739	&	IC 3284	&	Sa	&	Sab	&	Sab	&	Sab	&		\\
1996C$^\star$	&	0	&	0.02752	&	MCG+08-25-047	&	Sa	&	Sb	&		&	Sb	&		\\
2006ev	&	0	&	0.02762	&	UGC 11758	&	S	&	Sbc	&	S?	&	Sbc	&		\\
2005eq	&	0	&	0.02788	&	MCG-01-09-006	&	Sbc	&	S?	&	Scd?	&	Sbc	&		\\
2003U	&	0	&	0.02818	&	UGC 10832	&	Sc	&	Sc	&	Scd:	&	Sc	&		\\
2002de	&	0	&	0.02827	&	NGC 6104	&	S0	&	Sb	&	S?	&	Sb	&		\\
2009ad	&	0	&	0.02834	&	UGC 3236	&	Sbc	&	Sb	&	Sbc	&	Sbc	&		\\
2006qo	&	0	&	0.02885	&	UGC 4133	&	Sc	&	Sc	&	Scd:	&	Sc	&		\\
2003ch	&	0	&	0.02922	&	UGC 3787	&	E/S0	&	E-S0	&	S0?	&	E/S0	&		\\
1994Q	&	0	&	0.02956	&	2MASX J16495110+4025599	&	S/Irr	&	Sc	&	Scd	&	Sc	&		\\
2007is	&	0	&	0.02968	&	UGC 10553	&	S/Irr	&	Sab	&	Sab:	&	Sab	&		\\
2004ef	&	0	&	0.02979	&	UGC 12158	&	Sb	&	Sb	&	Sb	&	Sb	&		\\
2007cc	&	0	&	0.03002	&	2MASX J14084200-2135498	&	S...	&	Sc	&	Sc	&	Sc	&		\\
2002ck	&	0	&	0.03031	&	UGC 10030	&	Sb	&	Sab	&	Sb	&	Sab	&		\\
2007ux	&	0	&	0.03043	&	2MASX J10091969+1459268	&		&	S0-a	&		&	S0/a	&		\\
PTF10bjs	&	0	&	0.03052	&	MCG+09-21-083	&		&	Sb	&	Sb	&	Sb	&		\\
2006bw	&	0	&	0.03079	&	LEDA 1258718	&		&	E	&		&	E	&		\\
2006en	&	0	&	0.03080	&	MCG+05-54-41	&	Sc	&	Sc	&		&	Sc	&		\\
1999cc	&	0	&	0.03153	&	NGC 6038	&	Sbc	&	Sc	&	Sc	&	Sc	&		\\
2005lu	&	0	&	0.03154	&	MCG-03-07-40	&	Sd	&	Sbc	&	S.../Irr?	&	Sbc	&		\\
10026	&	1	&	0.03160	&	MCG+10-15-120	&	Sd	&	Sc	&		&	Sc	&		\\
2007bd	&	0	&	0.03185	&	UGC 4455	&	Sab	&	Sa	&	Sa	&	Sa	&		\\
2006te	&	0	&	0.03210	&	2MASX J08114347+4133184	&	Sbc	&	S?	&		&	Sbc	&		\\
2004as	&	0	&	0.03213	&	LEDA 1676859	&	S/I	&	Sd	&		&	Sd	&		\\
2007ob	&	0	&	0.03266	&	2MASX J23122598+1354503	&	S0	&	S0-a	&	S0	&	S0	&		\\
2008bw	&	0	&	0.03276	&	UGC 11241	&	Sb	&	Sb	&	Sb	&	Sb	&		\\
1997dg	&	0	&	0.03280	&	LEDA 5065169	&		&		&		&	Scd	&	\citet{Hakobyan+2012}	\\
2008gp	&	0	&	0.03285	&	MCG+00-09-074	&	Sb	&	Sa	&	Sa	&	Sa	&		\\
2005iq	&	0	&	0.03295	&	MCG-03-01-08	&	Sa	&	Sab	&	Sa	&	Sa	&		\\
2008gl	&	0	&	0.03297	&	UGC 881	&	E	&	E	&	E	&	E	&		\\
2004L	&	0	&	0.03341	&	MCG+03-27-38	&	Sb	&	Sc	&		&	Sc	&		\\
2006gr	&	0	&	0.03344	&	UGC 12071	&	Sb	&	Sb	&	Sb	&	Sb	&		\\
2003iv	&	0	&	0.03358	&	MCG+02-08-14	&	E...	&	E	&		&	E	&		\\
2008bq	&	0	&	0.03360	&	2MASX J06410310-3802083	&	Sa	&	Sab	&	Sa	&	Sa	&		\\
2003cq	&	0	&	0.03375	&	NGC 3978	&	S	&	Sb	&	Sbc:	&	Sb	&		\\
2003ae	&	0	&	0.03380	&	2MASX J09282257+2726402	&		&	S?	&		&	Sbc	&	\citet{2013AJ....146...86T}	\\
2008af	&	0	&	0.03411	&	UGC 9640	&	E	&	E	&	E	&	E	&		\\
2005be	&	0	&	0.03416	&	2MASX J14593310+1640070	&	Sa	&	S0-a	&		&	S0/a	&		\\
2002G	&	0	&	0.03449	&	MCG+06-29-043	&	Sa	&	E-S0	&		&	Sa	&		\\
1996bl	&	0	&	0.03481	&	2MASX J00361813+1123354	&		&		&		&	Sbc	&	\citet{Hakobyan+2012}	\\
2008at	&	0	&	0.03513	&	UGC 5645	&	Sb	&	Sb	&	Sb	&	Sb	&		\\
2007hu	&	0	&	0.03540	&	NGC 6261	&	Sa	&	S0-a	&	S0/a	&	S0/a	&		\\
2006mo	&	0	&	0.03597	&	MCG+06-02-17	&	S...	&	Sc	&	S?	&	Sc	&		\\
2008gb	&	0	&	0.03640	&	UGC 2427	&	Sbc	&	Sbc	&	Sb-c	&	Sbc	&		\\
2000cf	&	0	&	0.03646	&	MCG+11-19-25	&		&	Sbc	&		&	Sbc	&		\\
17784	&	2	&	0.03652	&	SDSS J032950.83+000316.0	&		&	Sc	&		&	Sc	&		\\
2007O	&	0	&	0.03659	&	UGC 9612	&	Sbc	&	Sc	&	Sc	&	Sc	&		\\
2002eu	&	0	&	0.03671	&	2MASX J01494273+3237303	&	S0/Sa	&		&		&	S0/a	&		\\
2006je	&	0	&	0.03712	&	2MASX J01505173+3305321	&	Sa	&	S0	&		&	S0	&		\\
2007cb	&	0	&	0.03753	&	2MASX J13581715-2322179	&	Sab	&	Sb	&	Sa-b	&	Sab	&		\\
2002bz	&	0	&	0.03762	&	MCG+05-34-033	&	dG	&	E	&	S?	&	S0	&		\\
1999ef	&	0	&	0.03799	&	UGC 607	&	Sc	&	Sc	&	Scd?	&	Sc	&		\\
2006ak	&	0	&	0.03890	&	2MASX J11093314+2837393	&		&	S0	&		&	Sab	&	\citet{2016AJ....152...50T}	\\
2008051	&	0	&	0.03908	&	SDSS J151958.87+045416.8	&		&		&		&	SF	&	\citet{2015ApJ...812...31J}	\\
2005lz	&	0	&	0.03917	&	UGC 1666	&		&		&		&	Sa	&	\citet{2016AJ....152...50T}	\\
2003fa	&	0	&	0.04016	&	MCG+07-36-033	&	Sb:...	&	Sb	&	S?	&	Sb	&		\\
2001az	&	0	&	0.04059	&	UGC 10483	&	S 	&	Sbc	&	S	&	Sbc	&		\\
2007kk	&	0	&	0.04119	&	UGC 2828	&	Sb	&	Sb	&	Sbc	&	Sb	&		\\
2005hf	&	0	&	0.04205	&	2MASX J01270614+1906587	&		&		&		&	Sa	&	\citet{Hakobyan+2012}	\\
2007nq	&	0	&	0.04243	&	UGC 595	&	E	&	E	&	S?	&	E	&		\\
2001ic	&	0	&	0.04296	&	NGC 7503	&	E...	&	E	&	E:	&	E	&		\\
2006gt	&	0	&	0.04362	&	2MASX J00561810-0137327	&		&		&		&	Sc	&	\citet{2013AJ....146...86T}	\\
10805	&	2	&	0.04397	&	2MASX J22594265-0000478	&	Sm/Im	&	E?	&		&	Ir	&		\\
2005hc	&	0	&	0.04497	&	MCG+00-06-03	&	E/S0	&	Sbc	&		&	Sab	&	\citet{2016AJ....152...50T}	\\
2008by	&	0	&	0.04584	&	SDSSJ120520.81+405644.4	&		&		&		&	SF	&	\citet{2015ApJ...812...31J}	\\
360156	&	1	&	0.04620	&	SDSS J100313.51+015343.2	&	S	&	Sc	&		&	Sc	&		\\
2004gu	&	0	&	0.04698	&	2MASX J12462478+1156577	&		&	S?	&		&	Sab	&	\citet{2016AJ....152...50T}	\\
2008050	&	0	&	0.04702	&	ULAS J133647.52+050829.6	&		&		&		&	SF	&	\citet{2013ApJ...770..107C}	\\
2006eq	&	0	&	0.04834	&	2MASX J21283758+0113490	&		&	E?	&		&	Sbc	&	\citet{2013AJ....146...86T}	\\
2006cq	&	0	&	0.04921	&	IC 4239	&	S...	&	S0-a	&	S?	&	S0/a	&		\\
530086	&	1	&	0.05020	&	LEDA 1153699	&		&	E-S0	&		&	E/S0	&		\\
1993ac	&	0	&	0.05021	&	LEDA 17787	&	E	&	E	&		&	E	&		\\
2006ah	&	0	&	0.05097	&	LEDA 994819	&		&		&		&	SF	&	\citet{2013ApJ...770..107C}	\\
2010dt	&	0	&	0.05294	&	2MASX J16431345+3240391	&		&	Sb	&	Sb	&	Sb	&		\\
2007ar	&	0	&	0.05335	&	MCG+10-19-62	&	S0	&	E-S0	&	E	&	S0	&		\\
2008ac	&	0	&	0.05351	&	LEDA 2317123	&		&	S?	&		&	Sc	&	\citet{Hakobyan+2012}	\\
1998dx	&	0	&	0.05389	&	UGC 11149 	&	Es...	&		&		&	E	&		\\
490007	&	1	&	0.05470	&	SDSS J121704.45+463737.0	&		&	S?	&		&	SF	&	\citet{aguado2018fifteenth}	\\
19968	&	2	&	0.05490	&	2MASX J01372378-0018422	&		&	E	&		&	E	&		\\
2003ic	&	0	&	0.05491	&	MCG-02-02-086	&	E	&	S0	&	S0	&	E	&		\\
2005hj	&	0	&	0.05592	&	LEDA 4131950	&		&		&		&	SF	&	\citet{Hakobyan+2012}	\\
2006py	&	0	&	0.05661	&	LEDA 3333560	&		&	E	&		&	E	&		\\
2006ob	&	0	&	0.05824	&	UGC 1333	&	Sa	&	Sb	&	Sb:	&	Sb	&		\\
2006oa	&	0	&	0.05884	&	LEDA 4019108	&		&		&		&	SF	&	\citet{Hakobyan+2012}	\\
2001ah	&	0	&	0.05891	&	UGC 6211	&	Sc	&	Sbc	&	Sbc	&	Sbc	&		\\
2008bz	&	0	&	0.06143	&	2MASX J12385810+1107502	&		&	Sc	&		&	Sc	&		\\
2007ae	&	0	&	0.06416	&	UGC 10704	&	S	&	Sbc	&	S	&	Sbc	&		\\
10028	&	2	&	0.06426	&	2MASX J01105805+0016343	&	E/S0	&	E	&		&	E/S0	&		\\
6057	&	2	&	0.06651	&	LEDA 1130011	&	Sm/Im	&	S?	&		&	Sb	&	\citet{Hakobyan+2012}	\\
2006cj	&	0	&	0.06839	&	2MASX J12592407+2820498	&		&	S?	&		&	Sb	&	\citet{Hakobyan+2012}	\\
2006on	&	0	&	0.06884	&	LEDA 4524675	&		&	E?	&		&	E	&		\\
2006al	&	0	&	0.06905	&	LEDA 3358371	&		&	E?	&		&	S0/a	&	\citet{2016AJ....152...50T}	\\
2008Y	&	0	&	0.07029	&	MCG+09-19-039	&		&	Sbc	&		&	Sbc	&		\\
17240	&	2	&	0.07153	&	SDSS J003434.00-011257.5	&		&	E	&		&	E	&		\\
2003hu	&	0	&	0.07472	&	2MASX J19113272+7753382	&		&		&		&	Sbc	&	\citet{Hakobyan+2012}	\\
7876	&	2	&	0.07489	&	LEDA 3116670	&	Sb	&	E?	&		&	Sb	&		\\
17186	&	2	&	0.07849	&	Anon J020627-0053	&		&		&		&	SF	&	\citet{Hakobyan+2012}	\\
12779	&	2	&	0.07891	&	LEDA 1188169	&	S	&	Sbc	&		&	Sbc	&		\\
12950	&	2	&	0.08141	&	SDSS J232640.14-005026.2	&		&	E?	&		&	SF	&	\citet{Hakobyan+2012}	\\
130308	&	1	&	0.08220	&	LEDA 2422566	&		&	S?	&		&	SF	&	\citet{2017AA...599A..71D}	\\
12781	&	2	&	0.08282	&	2MASX J00213789-0100383	&		&	E-S0	&		&	E/S0	&		\\
722	&	2	&	0.08504	&	2MASS J00024907+0045051	&	E/S0	&	E	&		&	E	&		\\
3592	&	2	&	0.08543	&	2MASX J01161269+0047265	&	Sb	&	Sa	&		&	Sb	&		\\
21502	&	2	&	0.08784	&	2MASX J23342408-0053250	&		&	E	&		&	E	&		\\
1241	&	2	&	0.08848	&	SDSS J223041.15-004634.5	&		&		&		&	Pa	&	\citet{2015ApJ...812...31J}	\\
590194	&	1	&	0.08960	&	SDSS J084056.87+443127.3	&		&		&		&	SF	&	\citet{aguado2018fifteenth}	\\
774	&	2	&	0.09227	&	SDSS J014151.28-005236.2	&	Sa	&	S?	&		&	Pa	&	\citet{2015ApJ...812...31J}	\\
18241	&	2	&	0.09391	&	SDSS J204933.00-004543.0	&		&		&		&	SF	&	\citet{2015ApJ...812...31J}	\\
2102	&	2	&	0.09401	&	SDSS J204853.04+001129.8	&	Sm/Im	&		&		&	Ir	&		\\
420100	&	1	&	0.09630	&	SDSS J221225.28+005104.8	&	S0	&	E	&		&	E	&		\\
10010	&	1	&	0.09940	&	SDSS J100325.83+010143.3	&		&		&		&	SF	&	\citet{2015ApJ...812...31J}	\\
10434	&	2	&	0.10288	&	2MFGC 16592	&	E/S0	&	Sc	&		&	E/S0	&		\\
13135	&	2	&	0.10337	&	SDSS J001641.85-002530.5	&		&	E-S0	&		&	E/S0	&		\\
20064	&	2	&	0.10351	&	2MASX J23542073-0055023	&		&	Sa	&		&	Sa	&		\\
18697	&	2	&	0.10638	&	SDSS J004453.81-005948.6	&		&		&		&	SF	&	\citet{Hakobyan+2012}	\\
20625	&	2	&	0.10683	&	SDSS J002243.95-002845.8	&		&	E	&		&	SF	&	\citet{Hakobyan+2012}	\\
500038	&	1	&	0.10720	&	COSMOS 2334037	&		&		&		&	SF	&	\citet{aguado2018fifteenth}	\\
21034	&	2	&	0.10750	&	SDSS J015234.16+011438.8	&		&	Sb	&		&	Sbc	&	\citet{Hakobyan+2012}	\\
7147	&	2	&	0.10886	&	SDSS J232004.44-000320.1	&	E/S0	&		&		&	E/S0	&		\\
20027	&	1	&	0.10980	&	SDSS J122520.40+460059.2	&		&	Sbc	&		&	Sbc	&		\\
18612	&	2	&	0.11364	&	SDSS J004909.12+003547.8	&		&	S0-a	&		&	S0/a	&		\\
8719	&	2	&	0.11628	&	SDSS J003053.26-004307.0	&	Sm/Im	&		&		&	Ir	&		\\
5395	&	2	&	0.11635	&	SDSS J031833.80+000724.0	&	Sbc/Sc	&		&		&	Sc	&		\\
2561	&	2	&	0.11741	&	2MASX J03052260+0051346	&	Sb	&	E	&		&	Sb	&		\\
16259	&	2	&	0.11771	&	LEDA 1177432	&		&	E	&		&	E	&		\\
1371	&	2	&	0.11797	&	SDSS J231729.69+002546.8	&	E/S0	&	E?	&		&	E/S0	&		\\
19953	&	2	&	0.12190	&	SDSS J221143.27+003445.5	&		&		&		&	SF	&	\citet{aguado2018fifteenth}	\\
18835	&	2	&	0.12262	&	SDSS J033444.49+002119.8	&		&		&		&	Pa	&	\citet{Hakobyan+2012}	\\
2916	&	2	&	0.12303	&	Anon J220341+0034	&		&		&		&	Sa	&	\citet{2008AJ....135.1766Z}	\\
16021	&	2	&	0.12336	&	SDSS J005522.52-002321.1	&		&		&		&	SF	&	\citet{aguado2018fifteenth}	\\
6406	&	2	&	0.12376	&	SDSS J030421.25-010347.1	&	Sb	&		&		&	Sb	&		\\
13044	&	2	&	0.12455	&	SDSS J221010.32+003014.1	&	Sc	&		&		&	Sc	&		\\
2992	&	2	&	0.12608	&	SDSS J034159.34-004658.4	&	Sb	&		&		&	Sab	&	\citet{2008AJ....135.1766Z}	\\
16069	&	2	&	0.12688	&	SDSS J224458.81-010022.9	&		&		&		&	SF	&	\citet{Hakobyan+2012}	\\
744	&	2	&	0.12694	&	SDSS J215647.64+001901.3	&	Sm/Im	&		&		&	SF	&	\citet{aguado2018fifteenth}	\\
18855	&	2	&	0.12715	&	SDSS J031432.11+001608.0	&		&		&		&	SF	&	\citet{Hakobyan+2012}	\\
18809	&	2	&	0.12837	&	SDSS J032331.35+004002.1	&		&		&		&	Pa	&	\citet{Hakobyan+2012}	\\
22075	&	2	&	0.12899	&	SDSS J015951.28+011259.7 	&		&		&		&	Pa	&	\citet{Hakobyan+2012}	\\
1032	&	2	&	0.12903	&	SDSS J030711.01+010711.9	&	Sa	&		&		&	Sa	&		\\
5751	&	2	&	0.12928	&	SDSS J004632.24+005017.3	&	Sbc/Sc	&		&		&	Sbc	&		\\
17280	&	2	&	0.13045	&	SDSS J034310.04+000614.2	&		&		&		&	Sbc	&	\citet{Hakobyan+2012}	\\
15508	&	2	&	0.13353	&	SDSS J014840.67-003432.7	&		&		&		&	SF	&	\citet{aguado2018fifteenth}	\\
15234	&	2	&	0.13514	&	SDSS J010749.93+004942.9	&		&		&		&	SF	&	\citet{Hakobyan+2012}	\\
17629	&	2	&	0.13639	&	SDSS J020232.75-010523.7	&	S	&		&		&	SF	&	\citet{aguado2018fifteenth}	\\
18602	&	2	&	0.13696	&	SDSS J223556.07+003632.7	&		&		&		&	SF	&	\citet{aguado2018fifteenth}	\\
21062	&	2	&	0.13729	&	SDSS J221343.61+002346.6	&		&		&		&	SF	&	\citet{aguado2018fifteenth}	\\
17366	&	2	&	0.13811	&	SDSS J210308.39-010152.2	&		&		&		&	SF	&	\citet{Hakobyan+2012}	\\
190340	&	1	&	0.13840	&	SDSS J221632.39+002824.3	&		&		&		&	SF	&	\citet{aguado2018fifteenth}	\\
1794	&	2	&	0.14070	&	SDSS J211120.86-002643.4	&	Sm/Im	&		&		&	Ir	&		\\
2635	&	2	&	0.14310	&	SDSS J033048.96-011415.4	&	Sbc/Sc	&		&		&	Sc	&		\\
17497	&	2	&	0.14387	&	SDSS J022832.76-010234.1	&		&		&		&	SF	&	\citet{Hakobyan+2012}	\\
8921	&	2	&	0.14409	&	SDSS J214000.47-000029.0	&	Sm/Im	&		&		&	Ir	&		\\
17605	&	2	&	0.14533	&	SDSS J203648.61+000554.6	&		&		&		&	SF	&	\citet{aguado2018fifteenth}	\\
2031	&	2	&	0.15186	&	SDSS J204810.43-011016.8	&	Sm/Im	&		&		&	Ir	&		\\
19353	&	2	&	0.15325	&	SDSS J025227.18+001506.2	&		&		&		&	SF	&	\citet{Hakobyan+2012}	\\
5550	&	2	&	0.15492	&	SDSS J001423.63+001959.4	&		&		&		&	Sb	&		\\
18030	&	2	&	0.15517	&	SDSS J001943.97-002400.4	&		&		&		&	SF	&	\citet{Hakobyan+2012}	\\
13354	&	2	&	0.15653	&	SDSS J015015.53-005312.1	&		&		&		&	SF	&	\citet{Hakobyan+2012}	\\
17171	&	2	&	0.15899	&	SDSS J214600.83-011309.6	&		&		&		&	Pa	&	\citet{Hakobyan+2012}	\\
3317	&	2	&	0.15990	&	SDSS J014751.04+003825.5	&	Sm/Im	&		&		&	Ir	&		\\
2689	&	2	&	0.16035	&	SDSS J013936.00-004528.5	&		&		&		&	Pa	&	\citet{Hakobyan+2012}	\\
3087	&	2	&	0.16431	&	SDSS J012137.58-005837.7	&	Sm/Im	&		&		&	Ir	&		\\
19616	&	2	&	0.16455	&	SDSS J022823.91+001109.6	&		&		&		&	SF	&	\citet{Hakobyan+2012}	\\
20764	&	2	&	0.16477	&	SDSS J014428.99+001347.2	&		&		&		&	SF	&	\citet{aguado2018fifteenth}	\\
12843	&	2	&	0.16595	&	SDSS J213530.83-005846.6	&		&		&		&	Pa	&	\citet{Hakobyan+2012}	\\
12856	&	2	&	0.17028	&	SDSS J221127.68+004520.1	&		&		&		&	SF	&	\citet{Hakobyan+2012}	\\
3080	&	2	&	0.17315	&	SDSS J010743.60-010222.1	&	Sa	&		&		&	Sa	&		\\
15648	&	2	&	0.17383	&	SDSS J205452.51-001144.9	&		&		&		&	Pa	&	\citet{Hakobyan+2012}	\\
14421	&	2	&	0.17400	&	SDSS J020719.18+011507.2	&		&		&		&	Pa	&	\citet{Hakobyan+2012}	\\
19969	&	2	&	0.17428	&	SDSS J020738.36-001926.5	&		&		&		&	SF	&	\citet{Hakobyan+2012}	\\
5635	&	2	&	0.17839	&	SDSS J221243.88-000206.2	&	Sm/Im	&		&		&	Ir	&		\\
6936	&	2	&	0.17890	&	SDSS J213256.13-004200.2	&	Sm/Im	&		&		&	Ir	&		\\
2372	&	2	&	0.17958	&	SDSS J024205.00-003227.7	&	E/S0	&		&		&	E/S0	&		\\
13254	&	2	&	0.17990	&	SDSS J024814.09-002048.5	&		&		&		&	SF	&	\citet{Hakobyan+2012}	\\
14284	&	2	&	0.18037	&	SDSS J031611.84-003603.5	&		&		&		&	Pa	&	\citet{Hakobyan+2012}	\\
17215	&	2	&	0.18079	&	LEDA 1184310	&		&		&		&	Sab	&	\citet{Hakobyan+2012}	\\
15443	&	2	&	0.18123	&	SDSS J031928.18-001904.8	&		&		&		&	SF	&	\citet{Hakobyan+2012}	\\
15421	&	2	&	0.18443	&	SDSS J021457.91+003609.7	&		&		&		&	SF	&	\citet{Hakobyan+2012}	\\
8213	&	2	&	0.18468	&	SDSS J235005.06-005517.5	&	Sbc/Sc	&		&		&	Sbc	&		\\
6304	&	2	&	0.18979	&	SDSS J014559.74+011144.4	&	Sm/Im	&		&		&	Ir	&		\\
762	&	2	&	0.19009	&	SDSS J010208.65-005246.7	&	Sa	&		&		&	Sa	&		\\
2246	&	2	&	0.19422	&	SDSS J032021.71-005305.3	&	Sm/Im	&		&		&	Ir	&		\\
16099	&	2	&	0.19580	&	SDSS J014541.09-010316.5	&		&		&		&	SF	&	\citet{Hakobyan+2012}	\\
15129	&	2	&	0.19611	&	SDSS J211536.49-001918.1	&		&		&		&	SF	&	\citet{Hakobyan+2012}	\\
13070	&	2	&	0.19718	&	SDSS J235108.37-004447.6	&		&		&		&	SF	&	\citet{Hakobyan+2012}	\\
15222	&	2	&	0.19801	&	SDSS J001124.57+004207.2	&		&		&		&	E/S0	&	\citet{Hakobyan+2012}	\\
7243	&	2	&	0.20323	&	SDSS J215219.02+002818.9	&	Sm/Im	&		&		&	Ir	&		\\
17801	&	2	&	0.20515	&	SDSS J210422.51-005354.4	&		&		&		&	Sb	&	\citet{Hakobyan+2012}	\\
19913	&	2	&	0.20557	&	SDSS J221502.93-002030.1	&	S0/a	&		&		&	S0/a	&		\\
7847	&	2	&	0.21160	&	SDSS J020950.32-000342.1	&	Sb	&		&		&	Sb	&		\\
2330	&	2	&	0.21179	&	SDSS J002713.76+010715.0	&	Sb	&		&		&	Sb	&		\\
8495	&	2	&	0.21353	&	SDSS J222102.64-004454.2	&	Sb	&		&		&	Sb	&		\\
9467	&	2	&	0.21885	&	SDSS J215548.23+011052.6	&	Sa	&		&		&	Sa	&		\\
5533	&	2	&	0.21887	&	SDSS J215440.79+002446.0	&	Sm/Im	&		&		&	Ir	&		\\
13072	&	2	&	0.22916	&	SDSS J221950.56+000125.2	&		&		&		&	SF	&	\citet{Hakobyan+2012}	\\
3452	&	2	&	0.22967	&	SDSS J221841.11+003822.2	&	Sm/Im	&		&		&	Ir	&		\\
12971	&	2	&	0.23380	&	SDSS J002635.42-001811.8	&		&		&		&	Pa	&	\citet{Hakobyan+2012}	\\
13511	&	2	&	0.23652	&	SDSS J024226.71-004739.2	&		&		&		&	Pa	&	\citet{Xavier2013}	\\
3377	&	2	&	0.24448	&	SDSS J033637.48+010443.7	&	Sm/Im	&		&		&	Ir	&		\\
3451	&	2	&	0.24835	&	SDSS J221616.45+004228.1	&	Sa	&		&		&	Sa	&		\\
15161	&	2	&	0.24852	&	SDSS J022322.22+004908.4	&		&		&		&	SF	&	\citet{Hakobyan+2012}	\\
3199	&	2	&	0.24961	&	SDSS J221309.91+010301.6	&	Sb	&		&		&	Sb	&		\\
5717	&	2	&	0.25037	&	SDSS J011135.04-000021.4	&	Sm/Im	&		&		&	Ir	&		\\
9032	&	2	&	0.25249	&	SDSS J223132.24-002937.1	&	Sm/Im	&		&		&	Ir	&		\\
1112	&	2	&	0.25609	&	SDSS J223604.05-002229.7	&	Sb	&		&		&	Sb	&		\\
9457	&	2	&	0.25672	&	SDSS J222315.51+001513.3	&	Sa	&		&		&	Sa	&		\\
8046	&	2	&	0.25760	&	SDSS J023628.25+003042.6 	&	E/S0	&		&		&	E/S0	&		\\
6108	&	2	&	0.25800	&	SDSS J000713.57+002056.7	&	Sm/Im	&		&		&	Ir	&		\\
2017	&	2	&	0.26162	&	SDSS J215546.53+003536.4	&	Sbc/Sc	&		&		&	Sbc	&		\\
1253	&	2	&	0.26166	&	SDSS J213511.66+000946.2	&	Sbc/Sc	&		&		&	Sbc	&		\\
2943	&	2	&	0.26405	&	Anon J011049+0100	&	Sm/Im	&		&		&	Ir	&		\\
13099	&	2	&	0.26451	&	SDSS J235916.47-011502.5	&		&		&		&	SF	&	\citet{Hakobyan+2012}	\\
6315	&	2	&	0.26576	&	SDSS J204155.82+010530.7	&	Sm/Im	&		&		&	Ir	&		\\
6192	&	2	&	0.27091	&	SDSS J231351.64+011526.2	&	Sm/Im	&		&		&	Ir	&		\\
4000	&	2	&	0.27656	&	SDSS J020404.01-002158.7	&	Sm/Im	&		&		&	Ir	&		\\
5957	&	2	&	0.27923	&	SDSS J021902.35-001621.2	&	Sm/Im	&		&		&	Ir	&		\\
6196	&	2	&	0.27980	&	SDSS J223031.48-003008.6	&	E/S0	&		&		&	E/S0	&		\\
2789	&	2	&	0.28890	&	SDSS J225648.48+002402.0	&	E/S0	&		&		&	E/S0	&		\\
6249	&	2	&	0.29353	&	SDSS J001303.75-003712.9	&	Sm/Im	&		&		&	Ir	&		\\
13610	&	2	&	0.29683	&	SDSS J214403.41+004331.7	&		&		&		&	SF	&	\citet{Hakobyan+2012}	\\
6137	&	2	&	0.29888	&	SDSS J203144.52+001441.8	&	Sbc/Sc	&		&		&	Sbc	&		\\
5391	&	2	&	0.30021	&	SDSS J032922.16-010542.9	&	Sm/Im	&		&		&	Ir	&		\\
6699	&	2	&	0.30915	&	SDSS J213115.63-010326.3	&	Sb	&		&		&	Sb	&		\\
5844	&	2	&	0.30929	&	SDSS J215108.58-005034.0	&	Sm/Im	&		&		&	Ir	&		\\
16211	&	2	&	0.30938	&	SDSS J231239.09+001557.5	&		&		&		&	Pa	&	\citet{Hakobyan+2012}	\\
4241	&	2	&	0.33051	&	SDSS J004857.01-005419.8	&	Sm/Im	&		&		&	Ir	&		\\
4679	&	2	&	0.33103	&	SDSS J012606.79+004036.9	&	Sm/Im	&		&		&	Ir	&		\\
05D3jr	&	3	&	0.37039	&	[HSP2005] J141928.768+525153.34	&	E	&		&		&	E	&		\\
7779	&	2	&	0.37986	&	SDSS J204019.15-000022.8	&	Sbc/Sc	&		&		&	Sbc	&		\\
18721	&	2	&	0.40127	&	SDSS J001218.66-000439.5	&		&		&		&	Pa	&	\citet{Hakobyan+2012}	\\
Vilas	&	4	&	0.93500	&		&		&		&		&	Early-type	&	\citet{2012ApJ...750....1M}	\\
Patuxent	&	4	&	0.97000	&		&		&		&		&	Late-type	&	\citet{2012ApJ...750....1M}	\\
Ombo	&	4	&	0.97520	&		&		&		&		&	Late-type	&	\citet{2012ApJ...750....1M}	\\
SCP05D0	&	4	&	1.01400	&		&		&		&		&	Early-type	&	\citet{2012ApJ...750....1M}	\\
Eagle	&	4	&	1.02000	&		&		&		&		&	Late-type	&	\citet{2012ApJ...750....1M}	\\
SCP06C0	&	4	&	1.09200	&		&		&		&		&	Early-type	&	\citet{2012ApJ...750....1M}	\\
Gabi	&	4	&	1.12000	&		&		&		&		&	Late-type	&	\citet{2012ApJ...750....1M}	\\
vespesian$^\dagger$	&	4	&	1.20600	&		&		&		&		&	E/S0	&	\citet{2014AJ....148...13R}	\\
Lancaster	&	4	&	1.23000	&		&		&		&		&	Early-type	&	\citet{2012ApJ...750....1M}	\\
Koekemoer	&	4	&	1.23000	&		&		&		&		&	Late-type	&	\citet{2012ApJ...750....1M}	\\
Aphrodite	&	4	&	1.30000	&		&		&		&		&	Late-type	&	\citet{2012ApJ...750....1M}	\\
Thoth	&	4	&	1.30500	&		&		&		&		&	Early-type	&	\citet{2012ApJ...750....1M}	\\
washington	&	4	&	1.33000	&	[RRS2014] GSD11Was Host G	&		&		&		&	Sb/Sbc/Sc	&	\citet{2014AJ....148...13R}	\\
Mcguire	&	4	&	1.37000	&		&		&		&		&	Late-type	&	\citet{2012ApJ...750....1M}	\\
Sasquatch	&	4	&	1.39000	&		&		&		&		&	Early-type	&	\citet{2012ApJ...750....1M}	\\
Primo	&	4	&	1.55000	&		&		&		&		&	Scd/Ir	&	\citet{2014AJ....148...13R}	\\
wilson	&	4	&	1.91400	&		&		&		&		&	E/S0	&	\citet{2014AJ....148...13R}	\\
colfax	&	4	&	2.26000	&		&		&		&		&	E/S0	&	\citet{2014AJ....148...13R}	\\
\hline
\end{longtable} 
$^\star$The coordinates for SN~1996C given in \textsc{Pantheon} ($\alpha=207.751587,~\delta=+49.341251$) are wrong. The correct coordinates are $\alpha=207.7025,~\delta=+49.318639$.

$^\dagger$The coordinates for the Hubble SN Vespesian given in \textsc{Pantheon} ($\alpha=215.136078,~\delta=+53.046726$) in fact correspond to another Hubble SN --- Obama. According to \citet{Riess+2018} the coordinates of SN Vespasian (CLF11Ves) are $\alpha=322.4275,~\delta=-7.696583$.
}